\documentclass{aastex}
\usepackage{emulateapj5}
\usepackage{rotate}
\usepackage{epsf}
\usepackage{onecolfloat}

\newcommand{\Msun}{{\rm M}_{\sun}}

\begin{document}
\twocolumn[

\title{Formation of globular clusters in hierarchical cosmology}

\author{Andrey V. Kravtsov}

\affil{Department of Astronomy and Astrophysics and 
       Kavli Institute for Cosmological Physics,\\
       The Enrico Fermi Institute,\\
       The University of Chicago, Chicago, IL 60637\\
       {\tt andrey@oddjob.uchicago.edu}}

\author{Oleg Y. Gnedin}

\affil{Space Telescope Science Institute\\
       3700 San Martin Drive, Baltimore, MD 21218\\
       {\tt ognedin@stsci.edu}}

\begin{abstract}  
  We study the formation of globular clusters in a Milky Way-size
  galaxy using a high-resolution cosmological simulation.  The
  clusters in our model form in the strongly baryon-dominated cores of
  supergiant molecular clouds in the gaseous disks of high-redshift
  galaxies.  The properties of clusters are estimated using a
  physically-motivated subgrid model of the isothermal cloud collapse.
  The first clusters in the simulation form at $z \approx 12$, while
  we conjecture that the best conditions for globular cluster
  formation appear to be at $z \sim 3-5$.  Most clusters form in the
  progenitor galaxies of the virial mass $M_{\rm h} > 10^9 \, \Msun$
  and the total mass of the cluster population is strongly correlated
  with the mass of its host galaxy: $M_{\rm GC} = 3 \times 10^6 \,
  \Msun (M_{\rm h}/10^{11} \, \Msun)^{1.1}$.  This corresponds to a
  fraction $\sim 2 \times 10^{-4}$ of the galactic baryons being in
  the form of globular clusters. In addition, the mass of the globular
  cluster population and the maximum cluster mass in a given region
  strongly correlate with the local average star formation rate. We
  find that the mass, size, and metallicity distributions of the
  globular cluster population identified in the simulation are
  remarkably similar to the corresponding distributions of the Milky
  Way globulars.  We find no clear mass-metallicity or age-metallicity
  correlations for the old clusters.  The zero-age mass function of
  globular clusters can be approximated by a power-law $dN/dM \propto
  M^{-\alpha}$ with $\alpha \approx 2$, in agreement with the mass
  function of young stellar clusters in starbursting galaxies.  We
  discuss in detail the origin and universality of the globular
  cluster mass function.  Our results indicate that globular clusters
  with properties similar to those of observed clusters can form
  naturally within dense gaseous disks at $z \gtrsim 3$ in the
  concordance $\Lambda$CDM cosmology.
\end{abstract}

]

\section{Introduction}
\label{sec:intro}

More than seventy years ago in a monograph entitled {\it Star
Clusters} Harlow \cite{shapley30} wrote: ``It is encouraging to see
how fragile and futile are the majority of astronomical theories and
speculations... for the futility of speculations emphasizes the
importance and durability of observations and indicates the steady
progress of the science.''  These words are particularly true for the
models of globular cluster (GC) formation.  Extensive observational
surveys of globular cluster systems in the Milky Way and other
galaxies have been compiled during the past two decades
\citep[e.g.,][]{harris01}.  At the same time, despite a wide variety
of proposed models, a self-consistent scenario of globular cluster
formation is yet to be constructed.

The existing models can be classified into four broad categories.
In the primary models globular clusters are envisioned to have formed
soon after recombination, with masses determined by the cosmological
Jeans mass \citep{peebles_dicke68,peebles84}.  In the secondary
models, globular clusters are assumed to appear during the early
stages of galaxy formation.  For instance, \citet{fall_rees85} pointed
out that thermal instability in hot gaseous halos of young galaxies
can naturally lead to the condensation of globular cluster-size
clouds.  Several other trigger mechanisms operating during galaxy
formation, such as the shock compression and collisions of primordial
molecular clouds, were also explored \citep{gunn80,burkert_etal92,
murray_lin92,larson96,harris_pudritz94,cen01}.

Models of the third class correspond to the relatively recent stages
of galaxy formation.  \citet{schweizer87} and \citet{ashman_zepf92},
for example, proposed a model of GC formation in the gas-rich mergers
of disk galaxies.  There mergers perturb, compress, and shock the
interstellar medium which creates the very high pressure regions
conducive to GC formation.  Accordingly, this model predicted that young
clusters should be present in merging and interacting galaxies.  These
predictions were successfully confirmed by HST observations
\citep{whitmore_schweizer95,holtzman_etal96,whitmore_etal99,zepf_etal99}.

The division between the second and third classes of models is
somewhat blurred.  In the current hierarchical structure formation
paradigm, galaxy formation is a continuous process of merging and
accretion.  The fourth, most recent class of models thus incorporates
the elements of all previous classes within the hierarchical framework
\citep{cote_etal00,cote_etal02,beasley_etal02,gnedin03}.  Globular
clusters in these models are assumed to form in young disks which
undergo frequent mergers.  At this point, the models are largely
phenomenological and characterize the cluster formation using somewhat
ad hoc recipes, limiting their predictive power.  Nevertheless, they
have been fairly successful in reproducing many properties of
the observed GC populations.  The comparisons of various models to
observations can be found in \cite{harris01} and \cite{gnedin_etal01}.

The main obstacle to building a realistic and self-consistent physical
model of globular cluster formation has always been the uncertainty in
the initial conditions.  In fact, all of the models mentioned above
would produce star clusters in environments where the conditions agree
with the model assumptions.  Another unknown is the connection between
globular clusters and galaxies.  On the one hand, the largest globular
clusters have masses comparable to those of dwarf spheroidal galaxies
($\sim 10^7 \, M_{\sun}$).  On the other, globular clusters do not
seem to have extended dark matter halos \citep[e.g.,][]{moore96} and
in that respect differ fundamentally from galaxies.  There is also
significant disparity between the densities, velocity dispersions, and
structural parameters of dwarf galaxies and globular clusters
\citep{kormendy85}. In order to understand these differences we need a
self-consistent model which ties the formation of globular clusters to
the realistic formation and evolution of their parent galaxies.

The theory of hierarchical galaxy formation has matured in the last
decade, motivated by the theory of inflation, guided by observations,
and aided by elaborate numerical simulations.  In hierarchical
models, galaxies form via gravitational instability from 
small-amplitude initial Gaussian fluctuations with well-defined
statistical properties.  Recently, this scenario has been
spectacularly confirmed by the CMB anisotropy measurements and other
cosmological probes \citep[e.g.,][]{spergel_etal03}.  The spatially
flat cosmological model dominated by the dark energy and cold dark
matter ($\Lambda$CDM) favored by observations provides a solid
framework for the theory of globular cluster formation.

Cosmological simulations follow the hierarchical build-up of galaxies
self-consistently starting from the well-defined
initial conditions.  These simulations are
now reaching the level of sophistication and dynamic range sufficient
to study the formation and dynamics of giant molecular clouds in
galactic disks.  Therefore, we can address the formation of the
proto-cluster clouds without resorting to phenomenological
parameterization and directly study the details of when, where, and
how globular clusters formed.

The main goal of this work is to study the formation of globular
cluster populations in the hierarchical scenario using a very
high-resolution cosmological simulation.  Based on
observational evidence, we assume that clusters form in dense
isothermal cores of the super giant molecular clouds ubiquitous in
high-redshift galactic disks.  In addition, we use a simple model of
isothermal collapse to derive the properties of stellar clusters that
would form in such cores.  We then compare the derived properties of
model globular clusters with those of the GC population in the Galaxy
as well as with the populations of young GCs in external galaxies.

Many decades ago, Harlow Shapley used the distribution of
globular clusters to greatly expand and re-define the structure of our
Galaxy.  It is only fitting that we now apply our understanding of
galaxy formation to predict and explain the properties of globular
clusters.

\section{Numerical Simulations}

\subsection{Numerical techniques and Physical Processes}
\label{sec:numerics}

The simulation presented in this paper was performed using the
Eulerian gasdynamics$+N$-body Adaptive Refinement Tree (ART) code.
This code is based on the cell-based approach to adaptive mesh
refinement developed by \citet{khokhlov98}.  The algorithm uses
a combination of multi-level particle-mesh
\citep{kravtsov_etal97,kravtsov99} and shock-capturing Eulerian
methods \citep{vanleer79,colella_glaz85} to follow the evolution of
dark matter (DM) and gas, respectively. High dynamic range is achieved by
applying adaptive mesh refinement both to the gasdynamics and gravity
calculations.

Several physical processes critical to various aspects of galaxy
formation are implemented in this code: star formation; metal
enrichment and thermal feedback due to the supernovae type II and type
Ia (SNII/Ia); self-consistent advection of metals; metallicity- and
density-dependent cooling and UV heating due to the cosmological ionizing
background, using the cooling and heating rates tabulated in the
temperature range $10^2 < T < 10^9$~K for a grid of densities,
metallicities, and UV intensities using {\tt Cloudy}
\citep[ver. 96b4,][]{ferland_etal98}.  In the present simulations
we set a minimum temperature of $T_{\rm min}=300$~K.  The
cooling and heating rates take into account Compton heating/cooling of
plasma, UV heating, atomic and molecular cooling.  While the detailed
implementation of these processes is described elsewhere
\citep{kravtsov03a,kravtsov03}, below we summarize the details crucial
to this study.

We use a ``constant efficiency'' star formation prescription.  Namely,
the stars are formed with a {\em constant} timescale $\tau_{\ast}$ so
that the star formation rate is proportional to the local gas density,
$\dot{\rho}_{\ast}\propto \rho_{\rm g}$.  This prescription is
motivated by observations of the star forming regions
\citep[e.g.,][]{young_etal96,wong_blitz02} and appears to reproduce
the Schmidt-like law of star formation on kpc scales
\citep{kravtsov03}.  Star formation is allowed to take place only in
the coldest and densest regions, $T<T_{\rm SF}$ and $\rho_{\rm g} >
\rho_{\rm SF}$, but no other criteria (like the collapse condition
$\nabla\cdot {\bf v} < 0$) are imposed.  We use $\tau_{\ast}=4$~Gyr,
$T_{\rm SF}=9000$~K, and $\rho_{\rm SF}=1.64{\ \rm M_{\odot}pc^{-3}}$
or the atomic hydrogen number density of $n_{\rm H}=50{\ \rm
cm^{-3}}$.  The adopted values of $T_{\rm SF}$ and $\rho_{\rm SF}$ are
quite different from the typical temperatures and densities of star
forming molecular cores: $T\lesssim 30-50$~K and $n_{\rm H}\gtrsim
10^4{\ \rm cm^{-3}}$.  They are, however, more appropriate for the
identification of star forming regions on $\sim 100$~pc scales which
are resolved in the present simulation. In practice, $T_{\rm SF}$
is not relevant because most of the gas with $\rho>\rho_{\rm SF}$ is at
temperatures of just a few hundred degrees Kelvin.

Each newly formed stellar particle is treated as a single-age stellar
population and its feedback on the surrounding gas is implemented
accordingly. The word feedback is used here in a broad sense to
include the injection of energy and heavy elements (metals) via stellar
winds and supernovae, and the secular stellar mass loss.  Specifically,
we assume that the stellar initial mass function (IMF) is described by
the \citet{miller_scalo79} functional form with stellar masses in the
range $0.1-100\ \rm M_{\odot}$. All stars with $m_{\ast}>8{\ \rm
M_{\odot}}$ deposit $2\times 10^{51}$~ergs of thermal energy and a
mass $f_{\rm Z}m_{\ast}$ of heavy elements in their parent cell (no
delay of cooling is introduced in these cells). The metal fraction is
$f_{\rm Z}= {\rm min}(0.2,0.01 m_{\ast}-0.06)$, which crudely
approximates the results of \citet{woosley_weaver95}. In addition, the
stellar particles return a fraction of their mass and metals to the
surrounding gas at a secular rate $\dot{m}_{\rm
loss}=m_{\ast}\,\,C_0(t-t_{\rm birth} + T_0)^{-1}$ with $C_0=0.05$ and
$T_0=5$~Myr \citep{jungwiert_etal01}.  The released metals are
advected along with the gas.  The
code also accounts for SNIa feedback assuming a rate that slowly
increases with time and broadly peaks at the population age of
1~Gyr. We assume that the fraction $5\times 10^{-3}$ of mass in stars
between 3 and $8\ \rm M_{\odot}$ explodes as SNIa over the entire
population history and each SNIa dumps $2\times 10^{51}$ ergs of
thermal energy and ejects $1.3\ \rm M_{\odot}$ of metals into parent
cell. For the assumed IMF, 75 SNII (instantly) and 11 SNIa (over
several billion years) are produced by a $10^4\ \rm M_{\odot}$ stellar
population.

\subsection{Simulation Parameters}

The simulation we use in our analysis follows the early ($z\gtrsim 3$)
stages of evolution for a galaxy of typical mass: $\approx
10^{12}h^{-1}\ \rm M_{\odot}$ at $z=0$. At the analyzed epochs, the
galaxy has already built up a significant portion of its final mass:
$1.3\times 10^{10}h^{-1}\ \rm M_{\odot}$ at $z=9$ and $2\times
10^{11}h^{-1}\ \rm M_{\odot}$ at $z=4$.  The total galaxy mass,
$M_{\rm h}$, is defined as the mass enclosed within the radius of the
average density equal 340 times the mean matter density.  The
simulation starts from a random realization of the Gaussian density
field at $z=50$ in a periodic box of $6h^{-1}$ comoving Mpc with the power
spectrum \citep{hu_sugiyama96} appropriate to the flat $\Lambda$CDM
model: $\Omega_0=1-\Omega_{\Lambda}=0.3$, $\Omega_b=0.043$,
$h=H_0/100=0.7$, $n_s=1$, and $\sigma_8=0.9$. The parameters have
their usual meaning and are consistent with recent cosmological
constraints \citep[e.g.,][]{spergel_etal03}.

To increase mass resolution, a low resolution simulation was run
first and a galactic-mass halo was selected. A lagrangian region
corresponding to five virial radii of the object at $z=0$ was then
identified at $z=50$ and re-sampled with additional small-scale
waves \citep{klypin_etal01}.  The total number of DM particles in the
high-resolution lagrangian region is $2.64\times 10^6$ and each particle mass
is $m_{\rm DM}=9.18\times 10^5h^{-1}\ \rm M_{\odot}$. Outside the
high-resolution region the matter distribution was sampled with
$\approx 3\times 10^5$ higher mass particles.

As the matter distribution evolves, the code adaptively and
recursively refines the mesh in the high density regions.  Initially,
a uniform $64^3$ grid covered the entire computational box. The
lagrangian region, however, was always unconditionally refined to the
third refinement level, corresponding to an effective grid size of
$512^3$. Beyond the third level, a mesh cell was tagged for refinement
if its gas {\em or} DM mass exceeded 0.125 and 0.0625 times the mean
mass expected for the average density in each component in the zeroth
level (i.e., uniform grid) cell, respectively. The refinement follows
the collapse of $1.2\times 10^6h^{-1}\ \rm M_{\odot}$ (gas) and
$3.7\times 10^6h^{-1}\ \rm M_{\odot}$ (DM) mass elements in a
quasi-lagrangian fashion.  These masses can be loosely considered as
gas mass resolution until the maximum level is reached beyond which
refinements are not done.  In the run we use this level is set to
$l_{\rm max}=9$ and is reached by $z\approx 10$. Beyond this, the
notion of mass resolution for gas is not well defined because gas is
represented as a continuous medium on an Eulerian mesh. Once the maximum
refinement level is reached, the mass per
cell then is no longer constant but reflects the local gas density.
For example, cells of the 9th level have gas densities spanning the
range of more than six orders of magnitudeas the interstellar medium
is multi-phase with tenuous hot gas and very dense cold gas occupying
different regions (see, e.g., Fig.~\ref{fig:pdf} below).

The spatial resolution of the simulation is thus time-dependent.  As
the density increases, additional refiniment levels are added to keep
the mass per cell approximately constant.  The maximum allowed
refinement level $l_{\rm max}$ was set to nine and this level was
reached at $z\approx 10$.  In the simulation we present, the physical
size of the maximum refinement cell is $\sim 28$, $20$, $26$, $37$,
and $45h^{-1}$ pc at $z=12$, $9$, $6$, $4$, and $3$, respectively.
Thus, the change over the analyzed range of epochs is not very large.
A total of $\approx 1.1\times 10^7$ mesh cells was used at $z=4$ with
$\approx 2.5\times 10^5$ of them at refinement levels of 8 and 9.  The
high-density cold star forming disks within DM halos were refined to
$l_{\rm max}=9$.  The physical size of mesh cells was $\Delta
x_l=26.16\,\,[10/(1+z)] 2^{9-l}\ \rm pc$, where $l$ is the cell's
level of refinement.  Each refinement level was integrated with its
own time step $\Delta t_l=\Delta t_02^{-l}\approx 2^{9-l}\times 2
\times 10^4\ \rm yr$, where $\Delta t_0\lesssim 10^7\ \rm yr$ is the
global time step on the zeroth level set using the
Courant-Friedrichs-Levy condition.

\begin{figure*}[t]
\centerline{
\epsfysize=3.5truein  \epsffile{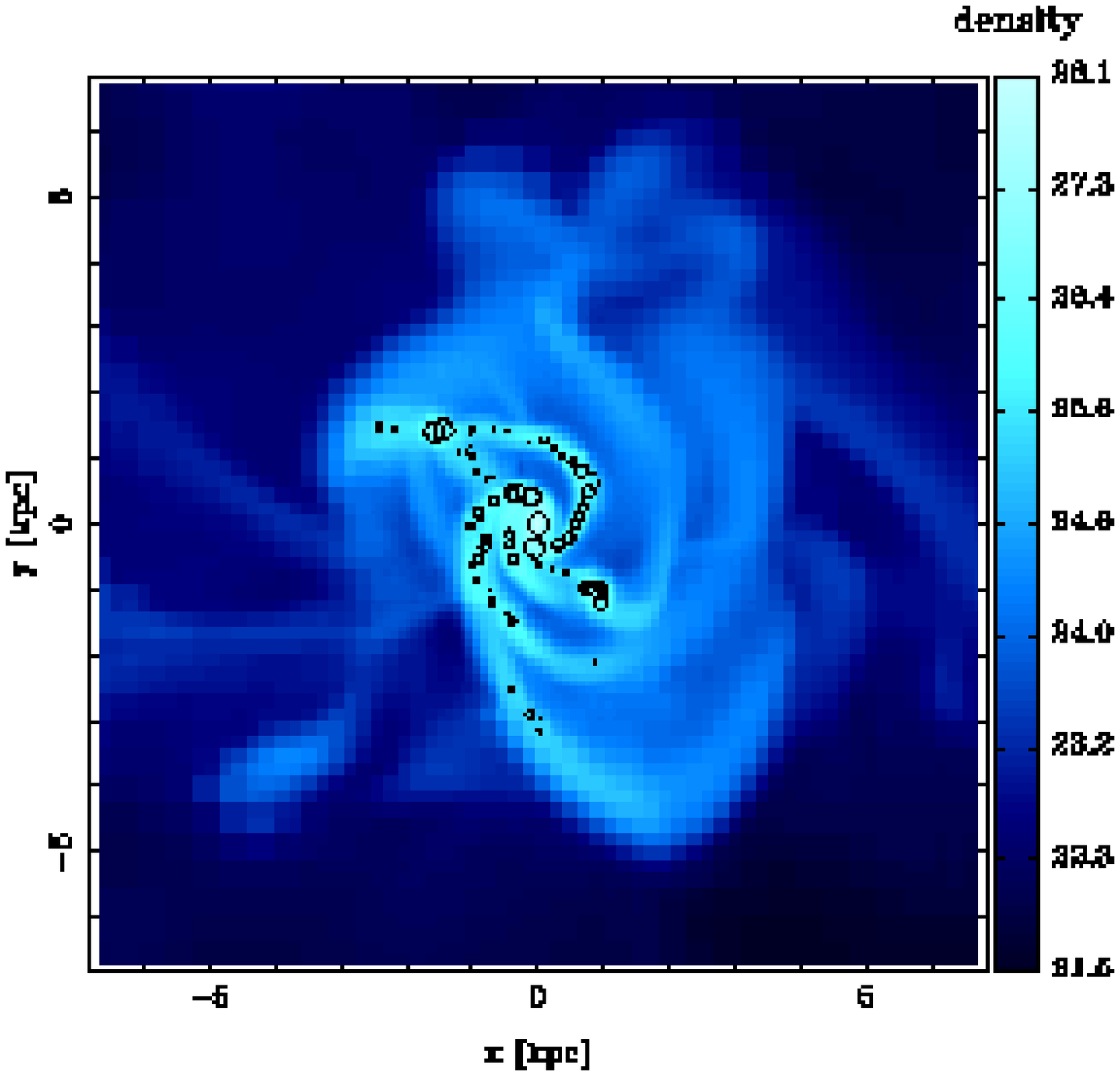}
 \epsfysize=3.5truein  \epsffile{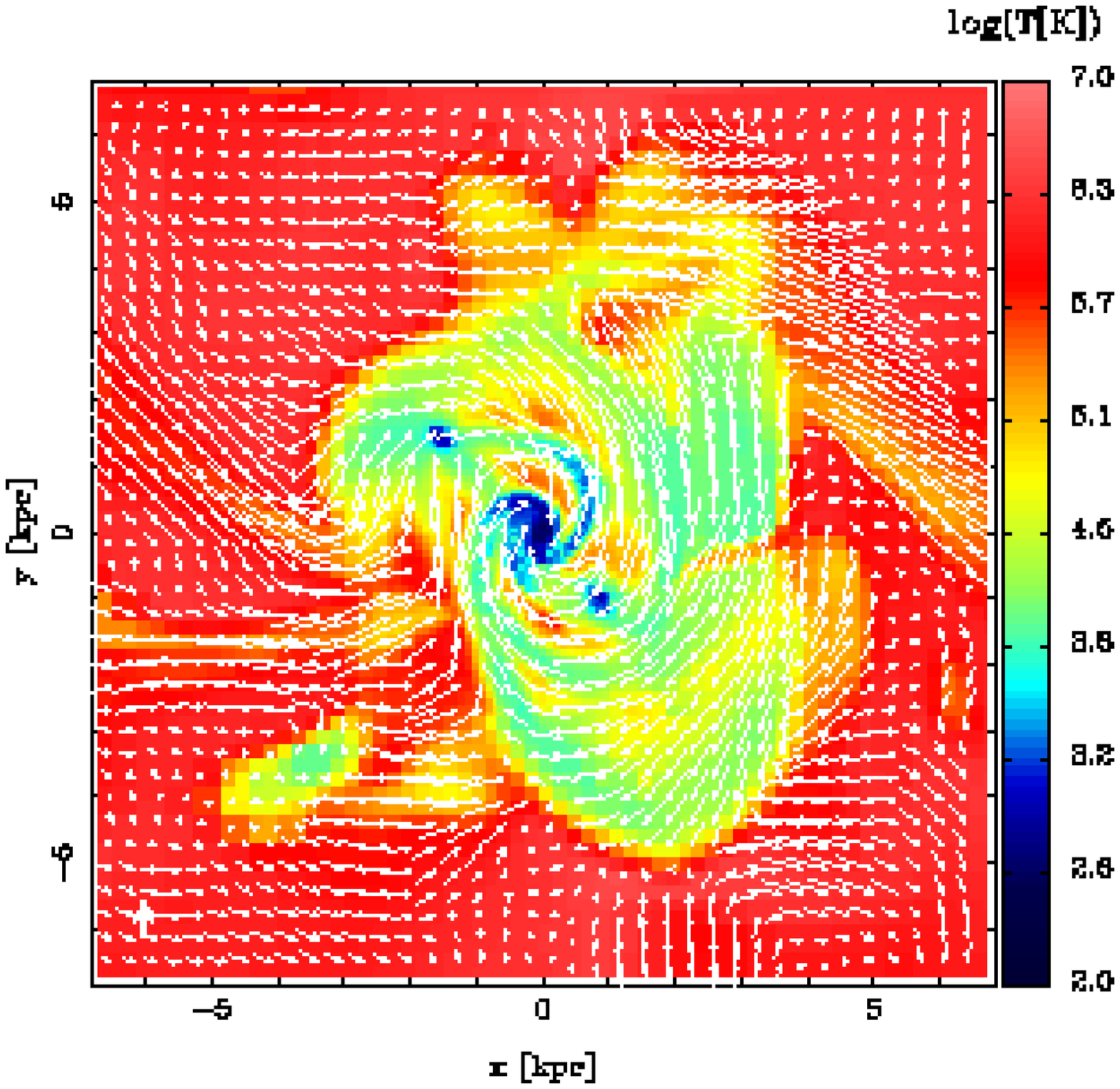}}
\caption{  Gas density (left panel) and temperature (right panel) in the most
  massive disk at $z=4$. The projected density is in
  cm$^{-2}$, while the density-weighted average temperature is in
  degrees Kelvin. The vectors in the right panel show gas velocities;
  the thick vertical vector in the lower left corner of the panel
  corresponds to $200\ {\rm km\,s^{-1}}$. The density and temperature
  are projected over a $3.5$~kpc slice centered on the cell of the
  maximum gas density (the center of the plot).  The figure shows a
  nearly face-on disk with prominent spiral arms in the process of
  very active accretion and merging. In our model the globular
  clusters form in the densest regions of the disk corresponding to
  the darkest knots in the temperature map. The globular clusters identified
  in the disk at this epoch are shown by circles in the left panel. The
radius of the circles corresponds to the mass of each cluster. 
  \label{fig:dtv}}
\end{figure*}

\subsection{Identification of the Globular Cluster Formation Sites}
\label{sec:formsites}

Although the resolution achieved in the simulation is very
high in the disk region, it is still insufficient to resolve the
formation of stellar clusters. The resolution, however, is sufficient
to identify the potential sites for GC formation.  The cores of giant
molecular clouds in high-redshift galaxies are the natural candidates
\citep{harris_pudritz94,mclaughlin_pudritz96} for such sites.
Numerical simulations of \cite{nakasato_etal00} show that globular
clusters with realistic masses and sizes can indeed form in such
cores.  We therefore adopt this picture and identify the cores of
dense gaseous clouds in simulated disks as the sites of globular
cluster formation.

We identify giant molecular clouds using the following algorithm. All
mesh cells with gas densities greater than a certain threshold
density, $\rho_{\rm mc}$, are selected and sorted into a list of
increasing density. The highest density cell and all of its immediate
neighbors are then included in the first cloud. This cell is labeled
as the core cell of the cloud.  The next highest density cell in the
list is then considered. If it happens to be already included in the
first cloud, all of its immediate neighbor cells are then also
included in the first cloud.  If, on the other hand, the cell is not
part of a cloud yet, it is assigned to a new cloud and is labeled as
its core.  The procedure repeats until the list of cells is exhausted.
The algorithm is thus somewhat similar, but not equivalent, to the
well-known friends-of-friends clustering algorithm.  The current
implementation will break a region into separate clouds for each
density peak rather than combining several peaks into the same cloud.

We explored several values of the threshold density $\rho_{\rm mc} =
1-50 \, \Msun$ pc$^{-3}$.  Although the cloud masses grow and ever
smaller clouds are included in the catalog as the threshold is
decreased, in the mass range relevant for the GC identification the
same cores are recovered for all $\rho_{\rm mc}$.  For our analysis
below we choose the cloud catalog with the fiducial value $\rho_{\rm
mc} = 1 \, \Msun$ pc$^{-3}$.  This corresponds to the gas number
density $\approx 40$ cm$^{-3}$ and pressure $\gtrsim 10^4$ K
cm$^{-3}$. Note that at these densities the gas temperature is at the
lowest value allowed in the simulation: $T_{\rm min} = 300$ K.

\subsection{The Subgrid Model}
  \label{sec:subgrid}

In order to derive the properties of star clusters forming in
molecular cloud cores, we complement the simulations with a
physical description of the gas distribution on a subgrid level.
Observations of star forming regions in the solar neighborhood
\citep[e.g.,][]{elmegreen02} show that clustered star formation
proceeds inevitably when the cores of molecular clouds
reach a critical density $\gtrsim 10^5$ cm$^{-3}$.  Formation of
young massive clusters is also accompanied by high external
pressure, $P > 10^7$ K cm$^{-3}$.  Within such cores, star
clusters form with a locally-high efficiency
\citep{geyer_burkert01,kroupa_etal01,kroupa_boily02}:
\begin{equation}
   \epsilon \equiv {M_{*} \over M_{\rm g}} \gtrsim 0.5,
\end{equation}
which is required in order to produce gravitationally-bound clusters.
On the theoretical side, analytical models and numerical simulations
also indicate that dense bound clusters can form quickly and
efficiently in the cores of giant molecular clouds of moderate
metallicity \citep[e.g.,][]{harris_pudritz94, mclaughlin_pudritz96,
nakasato_etal00}.  In the models of molecular clouds including thermal
support, turbulence, and magnetic fields star formation proceeds
rapidly, on one or two dynamical times \citep{pringle89,
ostriker_etal01, bate_etal03}.  Based on these observational and
theoretical results, we implement the following phenomenological
subgrid model.

At high densities in the central cells of the identified gas clouds
the cooling time is much shorter than the dynamical time, and
therefore the cells must be isothermal.  Indeed, the resolved
structure of the molecular clouds in the simulation has an isothermal
profile, $\rho_g \propto r^{-2}$.  We thus extrapolate this profile
inside the central unresolved cell.  (Such isothermal structure is
also predicted by the simulations of the collapse of cloud cores by
\citealt{nakasato_etal00}).  What we measure in the simulation is the
cell-averaged gas density, $\rho_{\rm cell} \equiv \rho_{\rm
av}(<R_{\rm cell})$, where $R_{\rm cell}$ is the cell radius (a half
of the cell dimension).  For an isothermal profile the average density
within $R_{\rm cell}$ is $\rho_{\rm av}(<R_{\rm cell}) = 3 \rho_{\rm
g}(R_{\rm cell})$.  We can thus derive the inner density profile as
\begin{equation}
   \rho_{\rm g}(r < R_{\rm cell}) = {1\over 3} \, \rho_{\rm cell} \,
                              \left(r \over R_{\rm cell}\right)^{-2}.
\end{equation}

We assume that a single cluster forms within the cloud core on a
dynamical (free-fall) time at the densities higher than the critical,
$\rho_{\rm csf}$.  To distinguish from the field star formation
denoted by the subscript 'SF' in \S\ref{sec:numerics}, here the
subscript 'csf' stands for 'clustered star formation'.  By choosing a
density threshold we postulate that only a high-density tail of the
gas distribution produces compact massive clusters, whereas the rest
(and most) of the gas participates in the formation of field stars or
open clusters.  This scenario is quite natural.  High gas densities
are required to match the observed stellar densities, which are
highest in globular clusters.  A single value of the density
threshold, of course, is not sufficient to describe complex physics of
molecular clouds but it can be very useful in defining the properties
of the clusters, as we demonstrate below.

The radius of the molecular core going into clustered star formation
$R_{\rm csf}$ is determined by the condition $\rho_{\rm g}(R_{\rm
csf}) = \rho_{\rm csf}$.  All gas within $R_{\rm csf}$ is converted
into stars with the efficiency $\epsilon$:
\begin{equation}
  M = \epsilon \, M_{\rm g}(R_{\rm csf}) 
    = \epsilon \, 4\pi \rho_{\rm csf} R_{\rm csf}^3.
  \label{eq:m}
\end{equation}
The fraction $1-\epsilon$ of the core mass remaining in the gas phase
will be expelled from that region, following the formation of
UV-bright O and B stars.  As a result of the gradual loss of the
remaining gas, the star cluster expands almost adiabatically
\citep[e.g.,][]{geyer_burkert01,boily_kroupa03} such that its final
size is
\begin{equation}
  R = {1 \over \epsilon} \, R_{\rm csf}
    = {1 \over \epsilon} \, R_{\rm cell} 
      \left({\rho_{\rm cell} \over 3\rho_{\rm csf}}\right)^{1/2}.
  \label{eq:r}
\end{equation}
The resulting average density of the clusters is
\begin{equation}
   \rho \equiv {M \over (4\pi/3) R^3}
      = \epsilon^4 {M_{\rm g}(R_{\rm csf}) \over (4\pi/3) R_{\rm csf}^3}
      = 3\epsilon^4 \rho_{\rm csf}.
  \label{eq:rho}
\end{equation}
We use the fiducial values $\epsilon = 0.6$ and $\rho_{\rm csf} = 10^4
\, \Msun$ pc$^{-3}$, which gives $\rho \approx 4 \times 10^3 \, \Msun$
pc$^{-3}$.  It is close to the median density at the half-mass radius
for the Milky Way globular clusters, which is $3 \times 10^3 \, \Msun$
pc$^{-3}$.

Although the expressions for the mass and size of star clusters are
linked to the resolution-dependent cell properties ($\rho_{\rm cell},
R_{\rm cell}$), the parameters of individual clusters are almost
insensitive to changes in the resolution.  This is because the cores
of dense molecular clouds in which globular clusters form are
isothermal.  Thus, when the cell size $R_{\rm cell}$ changes, the cell
density adjusts as $\rho_{\rm cell} \propto R_{\rm cell}^{-2}$,
leaving the radius $R_{\rm csf}$ and the cluster radius $R$ and mass
$M$ unchanged (see eqs.~[\ref{eq:m}] and [\ref{eq:r}]). This is indeed
what we find in the test presented below in \S~\ref{sec:convergence}.
When we repeat the analysis decreasing the level of refinement
($l_{\rm max}-1$), the masses of individual clusters do not change
significantly.

In the following, we present the properties of the model clusters
based on the density criterion. For completeness, in
\S\ref{sec:altern} we discuss and evaluate alternative subgrid models
and show that our model works best in reproducing the observed
properties of clusters.

Note that the mass function of globular clusters at birth will be
significantly modified by the effects of dynamical evolution.  As we
discuss in \S\ref{sec:future}, low-mass and low-density clusters are
preferentially dissolved by the combined effects of two-body
relaxation, tidal shocking, dynamical friction, and stellar evolution.
High-mass clusters ($M \gtrsim 10^5 \ \Msun$), on the other hand,
preserve their mass function and trace the initial distribution.

\begin{figure}[t]
\centerline{\epsfysize=3.4truein \epsffile{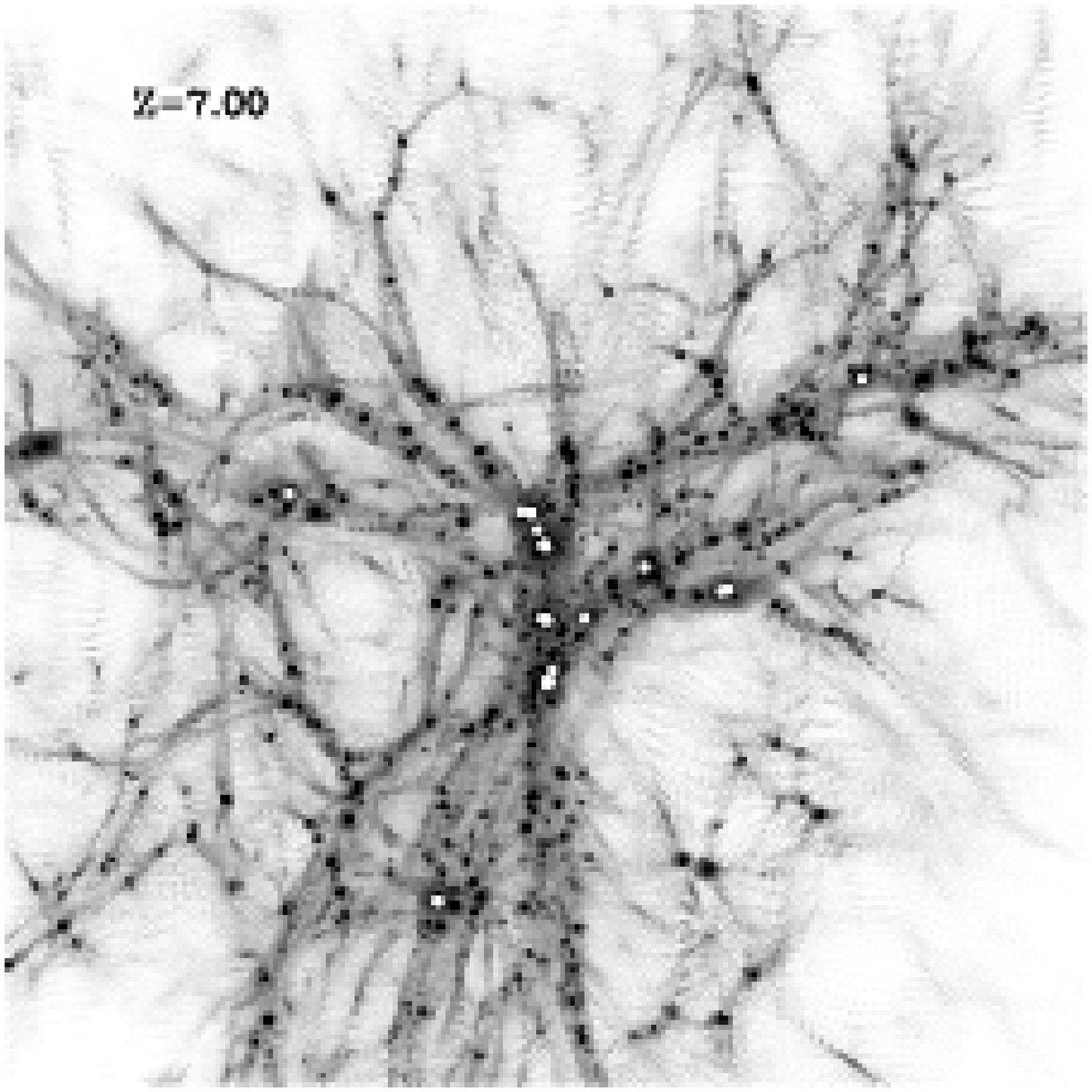}}
\centerline{\epsfysize=3.4truein \epsffile{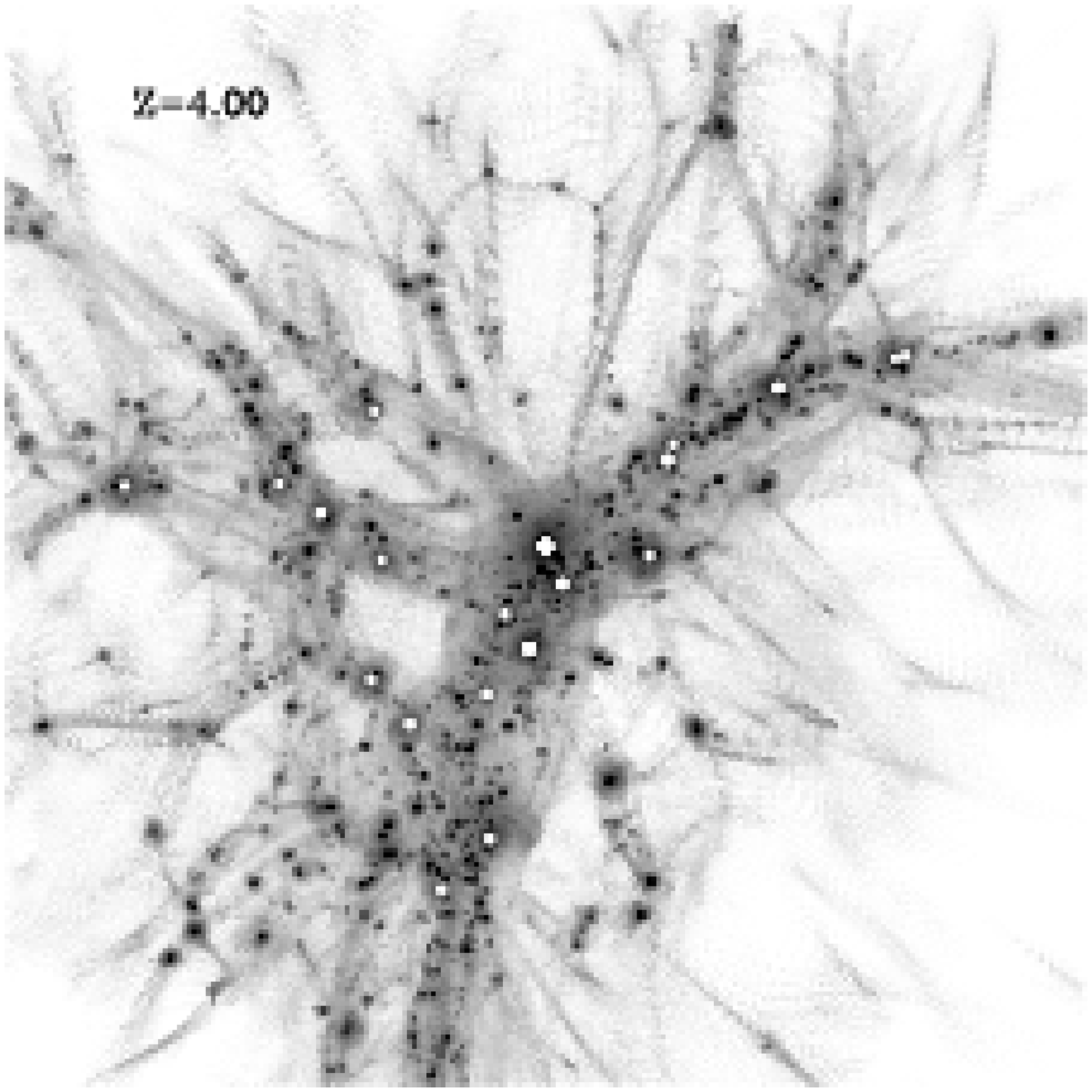}}
\caption{
  The identified globular clusters within the global distribution of
  dark matter at $z=7$ and $z=4$. The view is centered on the largest
  galaxy in the simulation and shows $1 h^{-1}$ Mpc region (comoving).
  The gray-scale colored particles represent the dark matter, while white
  circles in the centers of some halos show locations of the globular
  clusters identified in the simulation. Note that massive halos contain
  multiple clusters in their centers. The dark matter particles 
  are colored according to the local density on a logarithmic stretch. 
  \label{fig:dmgc}}
\end{figure}

\section{Results}
\label{sec:results}

We analyze the simulation outputs at twelve epochs between redshifts
$z = 11.8$ and $3.35$, identifying the cores of the giant molecular
clouds and computing properties of the model globular clusters, as
described in \S~\ref{sec:formsites} and \ref{sec:subgrid}.  The time
intervals between the outputs are in the range $\sim 100-300$~Myr.
Due to limited computational resources, the simulation was run only
until $z=3.3$.

\subsection{Spatial distribution of globular clusters at high redshifts}
\label{sec:spatial}

Before discussing the detailed properties of model clusters, we
first consider their spatial distribution.  Figure~\ref{fig:dtv} shows
the density and temperature maps, as well as the velocity field of
gas, in the region of the most massive disk at $z=4$.  The cooling of
the gas in these regions is very fast and the cold gas always settles
in a thin disk.  Frequent interactions drive strong spiral density
waves in the gaseous disks, which fragment into separate molecular clouds.
Globular clusters identified within those clouds are
represented by circles in the density map. The figure shows that
clusters in our model form in the dense cold regions, generally
tracing the spiral arms of the galaxy.  The morphology of the
distribution is very similar to that of young stellar clusters
observed in starburst galaxies
\citep[e.g.,][]{whitmore_schweizer95,zhang_etal01}.

At this epoch the parent halo of the disk experiences frequent mergers
and vigorous accretion of fresh gas.  The two cold knots outside the
spiral arms, for instance, are the small-mass satellite galaxies in
the process of merging with the central halo.  The dense gas in these
satellites could have been compressed by the external pressure and
shocks accompanying their collision with the disk.  The galaxy as a
whole exhibits frequent bursts of star formation associated with minor
and major mergers.  These mergers destroy the short-lived disks and
scatter away young stars and star clusters in a spheroidal halo.  The
cold gas, on the other hand, always falls back to form a new thin
disk.

Figure \ref{fig:dmgc} shows the spatial distribution of model globular
clusters within the large-scale structure formed in our simulation at
$z=7$ and $z=4$.  The distribution of dark matter is typical of
hierarchical models.  Visually, it is dominated by prominent filaments
on large scales and hundreds of dense dark matter halos tracing these
filaments on small scales.  The figure shows that parent halos of
globular clusters are tracing the skeleton of the large-scale
structure.  They concentrate in the densest regions of the filaments
close to the central massive object.  In other words, the distribution
of halos containing clusters is {\em highly biased} with respect to the
overall distribution of matter.  This bias is especially pronounced at
$z=7$.  The highly clustered distribution of halos at the early epochs
is a generic feature of the hierarchical models, in which objects of
galactic mass correspond to the relatively high peaks in the initial
Gaussian density field.  This property can be extremely important for
explaining the present distribution of globular clusters in the halo
of the Milky Way.  The high spatial bias of globular clusters at early
epochs would result in the more concentrated radial distribution of
globular clusters compared to the dark matter today \citep{west93} and
in the preferential location of higher-metallicity clusters towards to
the center, in agreement with observations
\citep{djorgovski_meylan94,vandenbergh03}. 

Note that although the high-redshift globular clusters form in dense
gaseous disks, the subsequent accretion of their parent galaxies along
filaments will lead to tidal stripping and disruption. For example,
analysis of the evolution of Milky Way size progenitors
\citep*{kravtsov_etal04} shows that most of the dwarf-size systems
that are located within $\lesssim 3$ virial radii from the progenitor
at $z>4$ accrete early and are disrupted before present day epoch.
This includes most of the objects hosting globular clusters in
Figure~\ref{fig:dmgc}. The disrupted systems form diffuse dark matter
halo and contribute to the stellar halo of the host
\citep*{bullock_etal01}. Their clusters would share the fate of the
stripped stars and should therefore have spatial distribution at $z=0$
similar to the stellar halo stars.  Direct observational evidence of
disruption is provided by the extended tidal tails around globular
clusters \citep[e.g.,][]{odenkirchen_etal03,sohn_etal03}, around dwarf
satellite galaxies \citep[][and references
therein]{freeman_bland-hawthorn02}, and the possible association
between the two \citep{lynden-bell95}.  We will investigate the
dynamical evolution and the present-day spatial distribution of the GC
population in our model in a future study.

\begin{figure}[t]
\plotone{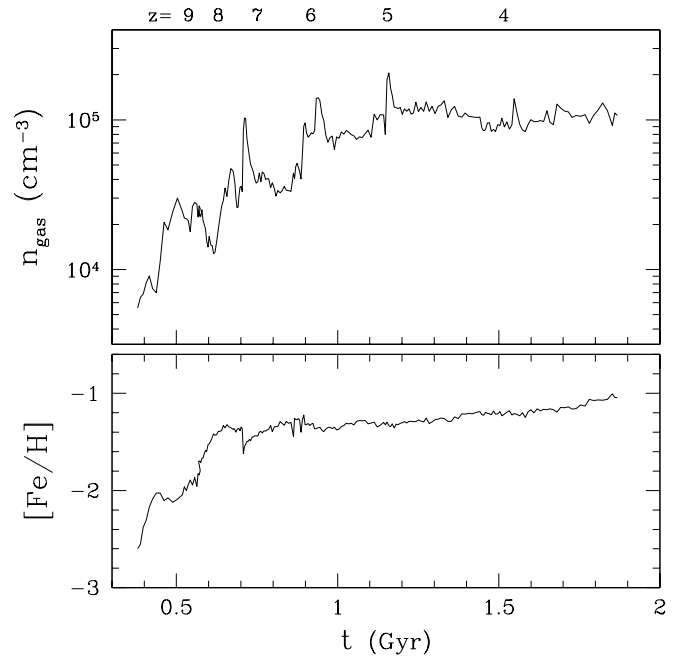}
\caption{Accretion history in the central cell of the main progenitor
  halo.  {\it Upper panel:} Gas number density.  The densities of dark
  matter and stars in this region are an order of magnitude smaller.
  {\it Lower panel:} Iron abundance of the gas with respect to the
  solar value.  The metallicity is due to SNII enrichment, as
  the contribution of SNIa is negligible at these epochs.
  \label{fig:all}}
\end{figure}

\subsection{Molecular clouds in the dense high-redshift disks}
\label{sec:GMC}

The globular clusters in our model form in the high-density cores of
giant molecular clouds of the high-redshift galaxies
(Fig.~\ref{fig:dtv}).  It is therefore important to consider the
properties of the molecular clouds in connection to the properties of
globular clusters and host galaxies.  Figure \ref{fig:all} shows the
evolution of density and metallicity in the central cell of the most
massive disk shown in Figure~\ref{fig:dtv}. Although this cell has the
highest gas density and is located at the bottom of potential well,
the overall evolution is common for all cells.  The gas density
exhibits several prominent peaks associated with the fast episodes of
accretion.  Cold metal-poor gas is delivered to the center of the
disk both by merging of smaller galaxies and by direct accretion of
gas along a filament that reaches inside the disk region.  The rapid
increase of the density and pressure in the cell during the accretion
events can trigger the collapse of the molecular cloud.  The gas density
saturates at lower redshifts ($z \lesssim 5$) as the accretion
on the center of the disk slows down.

The bottom panel of Figure \ref{fig:all} shows the evolution of
metallicity due to the SN ejecta (the contribution of SNIa to the
metallicity is negligible at these epochs).  The metallicity quickly
increases to about 10\% of solar and then evolves slowly.  Note that
the events of accretion of the fresh low-metallicity gas may lower the
mean metallicity, even in the very central region.  If a series of
globular clusters forms between $z=8$ and 4 in this region, the
younger ones are not necessarily more metal-rich than the older ones.
The age difference of these clusters would however be less than 2 Gyr.

Overall, the galaxies in the simulation exhibit a well-defined
correlation between the stellar mass and the average metallicity of
stars, $Z\propto M_{\ast}^{0.5}$, similar to the correlation observed
in nearby dwarf galaxies \citep{dekel_woo03}.  There is also a
significant spread in gas metallicity even within a single object,
which indicates that mixing of metals is rather inefficient.  The wide
range of gas metallicity in star forming regions eliminates any
clear age-metallicity correlation for high-redshift clusters.  For
instance, stars formed at the same epoch can have metallicities
different by up to two orders of magnitude.  This may also at least
partially explain the well-known ``second parameter problem''
\citep[e.g.,][]{carney01}.

The efficiency of GC formation, i.e. the ratio of the globular cluster
mass to the mass of the molecular, baryonic, or dark matter system
containing it, depends on the averaging scale.  Within the cores of
giant molecular clouds the local efficiency is of the order unity
(\S~\ref{sec:subgrid}).  Averaged over the whole molecular cloud
though, the efficiency is much lower because most of the molecular gas
is not participating in star formation at any given time.  When
compared with the total gas and/or dark matter mass in the galaxy, the
efficiency decreases by another order of magnitude.  There are thus
various types of globular cluster formation efficiencies, which we
consider in turn.

The detailed properties of the simulated molecular clouds depend on
the threshold density, $\rho_{\rm mc}$, used to define the cloud
boundary (see \S~\ref{sec:formsites}).  This boundary can be thought
of as an external tidal limitation.  The mass and size of the cloud
increase with the decreasing threshold density.  The cloud-scale
efficiency of globular cluster formation, which we define as
$\epsilon_{\rm GC} \equiv M/M_{\rm mc}$, varies accordingly.  We
find that the average efficiency is about $10^{-2}$ for $\rho_{\rm mc}
= 50 \, \Msun$ pc$^{-3}$, is in the range $10^{-3} - 10^{-2}$ for
$\rho_{\rm mc} = 10 \, \Msun$ pc$^{-3}$, and $10^{-4} - 10^{-3}$ for
$\rho_{\rm mc} = 1 \, \Msun$ pc$^{-3}$.  The estimated masses of
globular clusters, on the other hand, depend only the properties of
the cloud cores and are insensitive to the changes in the external
boundary condition.

For the massive globular clusters with $M > 3 \times 10^5 \, \Msun$
and the associated massive molecular clouds in our fiducial model with
$\rho_{\rm mc} = 1 \, \Msun$ pc$^{-3}$, the formation efficiency is
roughly constant: $\epsilon_{\rm GC} \approx 10^{-3}$.  However, if we
include lower mass clusters we find an anti-correlation with the cloud
mass (Spearman correlation coefficient $r_s = -0.35$).  In the range
$10^5 \, \Msun < M_{\rm mc} < 10^8 \, \Msun$, the relation is
$\lg{\epsilon_{\rm GC}} = -1.6 - (0.22 \pm 0.02)$ $\lg{(M_{\rm
mc}/\Msun)}$.  Overall, the numerical values of the efficiencies we
obtain are in good agreement with observations
\citep{harris_pudritz94}.

\begin{figure}[t]
\plotone{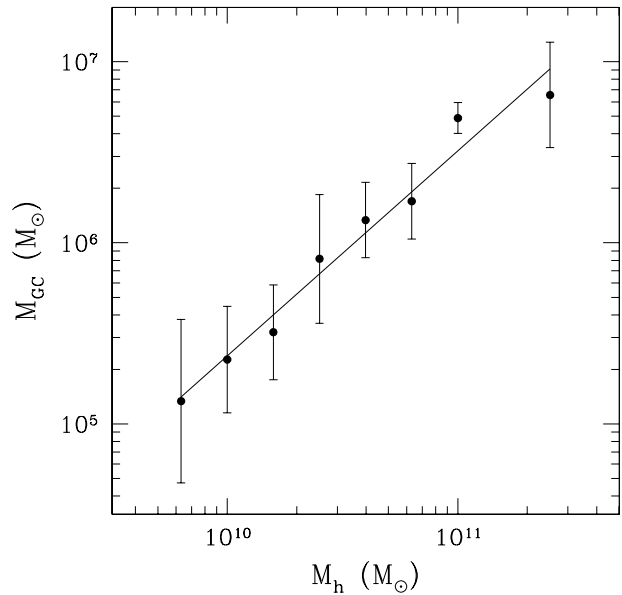}
\caption{The mass of the globular cluster system within a given halo vs. 
  the total mass of its parent halo, combined over
  all analyzed epochs.  Dots show the average in bins of
  width $\Delta\lg{M_{\rm h}} = 0.2$, while the error bars show the $1\sigma$
  deviations within the bin (not the error of the mean).  The solid line
  is the least-squares fit with the slope $d\lg{M_{\rm GC}}/d\lg{M_{\rm h}}
  = 1.13 \pm 0.08$.
  \label{fig:m_mh}}
\end{figure}

\subsection{The global efficiency of globular cluster formation}
  \label{sec:efficiency}
  
An important measure of the efficiency of globular cluster formation
is the total mass of clusters, $M_{\rm GC}$, within a parent galactic
halo.  For example, in giant elliptical galaxies the ratio of the
total cluster mass to the mass of stars plus the hot X-ray emitting
gas is roughly constant, $\varepsilon_{\rm GC}^{\rm b} \equiv M_{\rm
GC}/M_{\rm b} \approx 0.0026 \pm 0.0005$ \citep{mclaughlin99}.  This
parameter can be thought of as the efficiency of the conversion of
baryons into globular clusters.  In massive objects the baryon mass
$M_{\rm b}$ relates to the total galaxy mass $M_{\rm h}$ via the
universal baryon fraction, $M_{\rm b}/M_{\rm h} \approx f_{\rm b}
\approx 0.14$.  Thus perhaps even more fundamental is the ratio of the
globular cluster mass to the total galaxy mass: $\varepsilon_{\rm
GC}^{\rm t} \equiv M_{\rm GC}/M_{\rm h}$.

Figure \ref{fig:m_mh} shows the sum of the globular cluster masses in
each halo versus the progenitor galaxy mass at the time of GC
formation.  There is a well-defined correlation of the form
\begin{equation}
  M_{\rm GC} = 3.2 \times 10^6 \, \Msun
     \left({M_{\rm h} \over 10^{11} \, \Msun}\right)^{1.13 \pm 0.08}
  \label{eq:msum}
\end{equation}
albeit with scatter.  The global efficiency $\varepsilon_{\rm
GC}^{\rm t}$ is therefore only weakly dependent on the galaxy mass.
For most halos harboring massive clusters we find $\varepsilon_{\rm
GC}^{\rm t} = (2-5) \times 10^{-5}$.  The global baryon efficiency is
in the range $\varepsilon_{\rm GC}^{\rm b} = (2-3) \times 10^{-4}$,
and it scales with the galaxy mass as
\begin{equation}
  {M_{\rm GC} \over M_{\rm b}} = 2.5 \times 10^{-4}
     \left({M_{\rm h} \over 10^{11} \, \Msun}\right)^{0.25 \pm 0.12}.
  \label{eq:epsb}
\end{equation}

While these values are lower than those found by \citet{mclaughlin99},
they are in fact appropriate for a spiral galaxy like our own.
\citet{mclaughlin99} derived $\varepsilon_{\rm GC}^{\rm b} = 0.0027$
for the Galaxy taking into account only the mass of the stellar
spheroid.  We are interested here in scaling with the total mass and
therefore take the most recent estimate of the virial mass of the
Milky Way $M_{\rm h} \approx 10^{12} \, \Msun$ \citep{klypin_etal02}.
The mass of the observed globular clusters is $M_{\rm GC} = 5.2 \times
10^7 \, \Msun$ \citep{harris96}, and as \citet{mclaughlin99} argues,
this mass cannot differ from the initial mass by more than 25\%.  Thus
the global efficiency for the Galaxy is $\varepsilon_{\rm GC}^{\rm t}
\gtrsim 5 \times 10^{-5}$, and the baryon efficiency $\varepsilon_{\rm
  GC}^{\rm b} \approx \varepsilon_{\rm GC}^{\rm t}/f_{\rm b} \approx 4
\times 10^{-4}$.  Both of these estimates agree with our derived
correlations, eqs. (\ref{eq:msum}) and (\ref{eq:epsb}).

The global baryon efficiency can be related to the commonly used
specific frequency, $S_N \equiv N_{GC} 10^{0.4 (M_V+15)}$.  Taking the
mean cluster mass, $2\times 10^5 \ \Msun$, and assuming a
mass-to-light ratio for old clusters, $M/L_V = 3$, we obtain $S_N =
1.2\times 10^3 \, M_{\rm GC}/M_*$.  The stellar galaxy mass $M_*$ is
not a very good proxy for the baryon mass $M_{\rm b}$ at high
redshifts, but it can be used for the comparison at low redshifts.
Thus, our efficiency $\varepsilon_{\rm GC}^{\rm b} \approx 3 \times
10^{-4}$ corresponds to $S_N \approx 0.4$, which is indeed observed
for Sc galaxies.

It is interesting also that the mass of the globular cluster
population and the maximum cluster mass in a given region strongly
correlate with the local average star formation rate density: $M_{\rm
max}\propto \Sigma_{\rm SFR}^{0.54\pm 0.07}$ and $M_{\rm GC}\propto
\Sigma_{\rm SFR}^{0.75\pm 0.06}$ at $z=3.3$, where the masses and star
formation rate were estimated taking into account clusters with
$M>5\times 10^4\ {\rm M_{\odot}}$ and stellar particles younger than
$5\times 10^7$~yr and averaging over the cells of 7.7 physical
kpc. Each averaging cell therefore represents a different progenitor
galaxy in the simulation. A similar correlation
 was reported for the observed nearby galaxies
\citep{larsen02}. The interpretation of this correlation is
straightforward in our model. The star formation rate depends
sensitively on the mass fraction of gas in cold high-density star
forming regions \citep{kravtsov03}. The massive clusters in our model
are also assumed to form in such regions.  Thus, both the star
formation rate and the mass of the globular cluster population are
controlled by the amount of gas in the densest regions of the ISM.

\begin{table*}[tb]
\tablenum{1}
\label{tab:gcc}
\caption{The quartiles of the size and metallicity distributions of the model and
         Galactic globular clusters}
\begin{center}
\small
\begin{tabular}{lcccccc}
\tableline\tableline\\
\multicolumn{1}{l}{        } & 
\multicolumn{1}{c}{$R_h(25\%)$}&
\multicolumn{1}{c}{$R_h(50\%)$} &
\multicolumn{1}{c}{$R_h(75\%)$} &
\multicolumn{1}{c}{[Fe/H](25\%)} & 
\multicolumn{1}{c}{[Fe/H](50\%)} & 
\multicolumn{1}{c}{[Fe/H](75\%)} 
\\
\\
\tableline
\\
$z = 10$          & 1.9 & 2.0 & 2.3 & -2.7 & -2.6 & -2.5 \\
$z = 7$           & 2.0 & 2.4 & 2.7 & -2.2 & -1.8 & -1.6 \\
$z = 4$           & 2.0 & 2.4 & 2.9 & -1.8 & -1.5 & -1.2 \\
$z = 3.3$         & 2.0 & 2.3 & 2.9 & -1.7 & -1.4 & -1.1 \\
\\
MW, [Fe/H]$<$-1   & 1.2 & 2.8 & 4.8 & -1.8 & -1.6 & -1.4 \\
MW, all           & 0.7 & 2.4 & 4.0 & -1.7 & -1.4 & -0.7 \\
\\
\tableline
\end{tabular}
\end{center}
\end{table*}

\begin{figure}[t]
\plotone{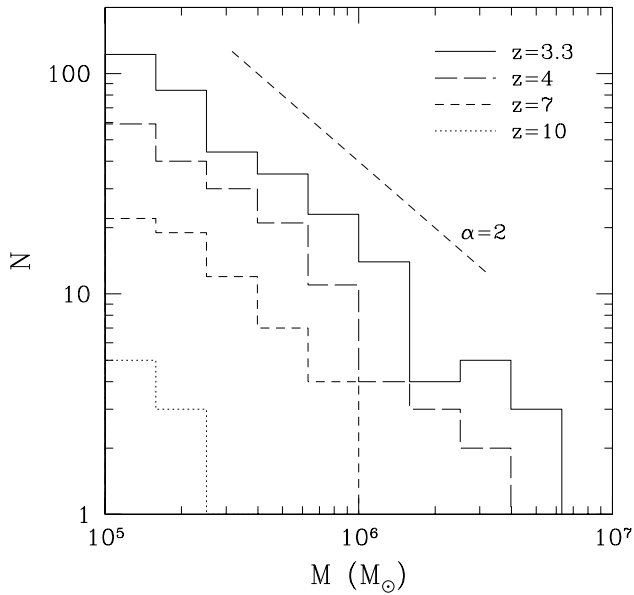}
\caption{The build-up of the initial mass function of globular clusters.
  Dotted, dashed, long-dashed, and solid histograms show cumulative
  distributions at $z = 10$, 7, 4, and 3.3, respectively.  The
  straight dashed line shows a power-law, $N\propto M^{-\alpha}$, with
  the slope $\alpha = 2$. Note that this is the mass function of
  young clusters, without accounting for the effects of dynamical evolution.
  \label{fig:gc_m}}
\end{figure}

\subsection{The mass, size, and metallicity distributions of the model clusters}

Figure \ref{fig:gc_m} shows the mass function of the model clusters
identified in all output epochs prior to a given redshift.  At all
epochs the distribution is well described by a power-law $dN/dM
\propto M^{-\alpha}$ at $M > 10^5 \, \Msun$.  The slope $\alpha$
evolves slowly and saturates at $\alpha = 2.05 \pm 0.07$ for $z \le
4$.  Note that  although the physical resolution of the
simulation changes somewhat with time, the subgrid prescription based
on the physical threshold density ensures that the cluster properties
are not affected.  In \S\ref{sec:altern} we show that the cluster mass
is a fraction of the mass of the central cell of the parent molecular
cloud that depends only on the physical density of the gas.  The mesh
in our simulation is refined in a quasi-lagrangian fashion, so as to
keep the same gas mass within a cell, and thus the cell masses and
cluster masses are always similar.

In this and subsequent figures we discard the most massive cluster
identified at the center of the most massive disk.  Such cluster would
be identified in observations as a compact galactic nucleus rather than
a distinct globular cluster.

\begin{figure}[t]
\plotone{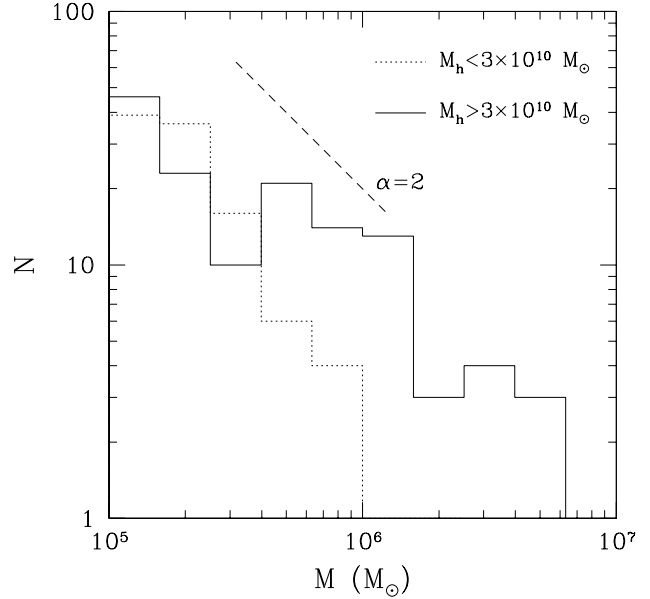}
\caption{The mass function of globular clusters formed within
  the parent halos of mass $<3\times 10^{10} \, \Msun$ ({\em dotted} histogram)
and $>3\times 10^{10} \, \Msun$ ({\em solid} histogram) at $z=3.3$.
  \label{fig:mhost}}
\end{figure}

The simplicity of the power-law shape of the mass function is
deceiving, as the mass functions of clusters in individual galaxies
exhibit a variety of shapes.  Figure \ref{fig:mhost} shows, for
example, that small halos form only clusters with $M < 10^6 \, \Msun$
with the mass function slope steeper than $\alpha = 2$, while large
halos form more massive clusters with a shallower mass function.
Remarkably, the convolution of the halo distribution with the
distribution of clusters within each halo produces the seemingly
invariant power-law mass function of globular clusters.  We
investigate the origin of the mass function in detail in
\S~\ref{sec:mf}.

In contrast to this power law function, the mass function of old
Galactic and most extragalactic globular clusters has traditionally
been described by a bell-shape (Gaussian) function.  The mass-to-light
ratio for old stars is approximately constant and it is appropriate to
use the luminosity function as a proxy to the mass function of
globular clusters.  Our results can be reconciled with observations
after taking into account the effects of subsequent dynamical
evolution of the model clusters.  Sophisticated models of the
dynamical evolution, started by Spitzer and collaborators in the 1970s
(see \citealt{spitzer87}) and refined in the 1990s
\citep[e.g.,][]{chernoff_weinberg90,gnedin_ostriker97,
murali_weinberg97,vesperini_heggie97,gnedin_etal99}, have shown that
tidally-truncated clusters undergo secular mass loss.  The main
processes shaping the GCMF are the evaporation through two-body
relaxation, stellar mass loss, tidal shocking by the host galaxy, and
stronger tidal truncation due to dynamical friction.
\citet{fall_zhang01} have demonstrated that these processes naturally
transform the initial power law into the observed truncated mass
function of old globular clusters.  Using a simple application of the
above results to our model clusters, we estimate that the dynamical
evolution will produce the peak and dispersion of the mass function of
surviving clusters in agreement with observations.  A more detailed
analysis, including the orbits of clusters in merging galaxies, will
be done in a subsequent study.

\begin{figure}[t]
\plotone{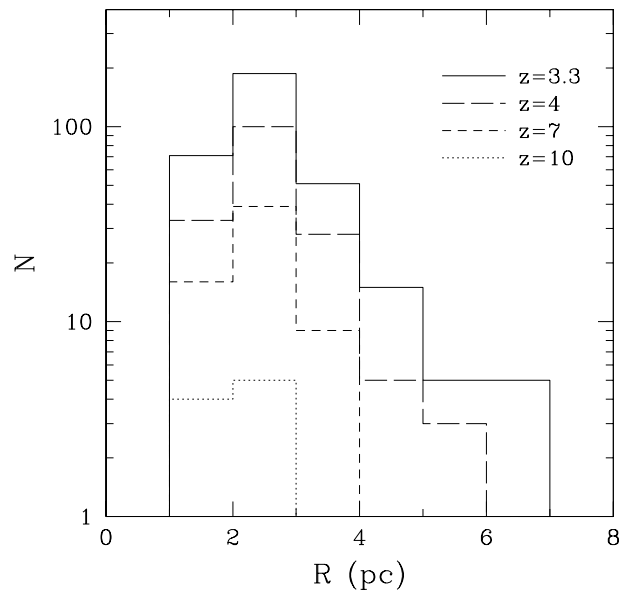}
\caption{The size distributions of globular clusters at four successive
  epochs, using only the massive clusters, $M > 10^5 \ \Msun$.
  \label{fig:gc_r}}
\end{figure}

Figure \ref{fig:gc_r} shows the distribution of the half-mass radii,
calculated according to our subgrid model (eq. [\ref{eq:r}]).  As are
the masses, the cluster sizes do not vary systematically with the
formation redshift and have consistent distributions at all epochs.
In this and the following figure we plot only the massive clusters, $M
> 10^5\ \Msun$, expected to survive the dynamical evolution.

The model size distribution is generally similar to the
observed sizes of the Galactic globular clusters
\citep[c.f.,][]{vandenbergh96}.  The differences at the smallest and
largest ends can be due to the following effects.  The adiabatic
expansion condition may not apply to the most massive clusters which
would then be larger.  The dynamical evolution effects, on the other
hand, would shrink the clusters and fill the range $R < 1$ pc.  Also,
some of the surviving clusters with $M < 10^5 \ \Msun$ may contribute
to the smallest size bin as well.

By construction, our constant density subgrid criterion leads to the
correlation between cluster masses and sizes, $R \propto M^{1/3}$.
Recent observations \citep{zepf_etal99,larsen04} suggest that the
sizes of young massive star clusters are almost independent of their
masses and show weaker correlation, $R \propto M^{0.1}$.  This implies
either that the massive star clusters form with intrinsically higher
densities or that the low-mass clusters expand more than we assumed
following the loss of the remaining gas in the star
forming cores.  For example, \cite{ashman_zepf01} suggested that the
formation efficiency $\epsilon$ increases with the binding energy of
the molecular cloud.  It might therefore be useful to define two
formation efficiencies: the mass conversion efficiency $\epsilon_M$ in
equation (\ref{eq:m}) and the expansion efficiency $\epsilon_R$ in
equation (\ref{eq:r}).  As we have argued in \S\ref{sec:subgrid},
$\epsilon_M$ cannot be much different from $0.5-0.6$, but $\epsilon_R$
can, in principle, scale with the cluster mass.  The observed trend
can be explained by $\epsilon_R \propto M^{0.2}$.  We do not discuss
these complications further in this paper but they should be addressed
in future work.

\begin{figure}[t]
\plotone{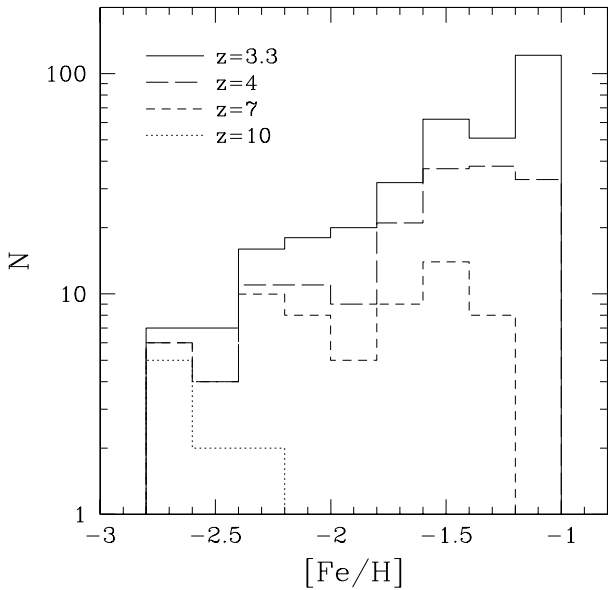}
\caption{The metallicity distributions of globular clusters at four
  successive epochs, using only the massive clusters, $M > 10^5 \ \Msun$.
  \label{fig:gc_z}}
\end{figure}

Figure \ref{fig:gc_z} shows the distribution of cluster metallicities,
which is remarkably similar to the metal-poor part of the Galactic
cluster distribution (see Table \ref{tab:gcc}).  Interestingly, we do
not find any correlation between the mass and metallicity of globular
clusters.  The Spearman rank correlation coefficient for the
cumulative distribution at $z \approx 3.3$ is $r_s = -0.06$, which is
consistent with no correlation.  Similarly, there is no correlation for
the Galactic GCs \citep{djorgovski_meylan94}.  The absence of the
mass-metallicity correlation may be due to a wide range of gas
metallicities in the star forming regions.  As we mentioned above,
this also explains the lack of a well-defined age-metallicity
correlation.

At the high redshifts considered here, most of the metals are
contributed by SNII which underproduce iron compared to the
$\alpha$-peak elements.  In order to calculate the fraction of ejected
metals contributed by iron, $\eta_{\rm Fe}$, we use the iron yields of
\cite{woosley_weaver95}, integrated over the \citet{miller_scalo79}
IMF in the range 12 to 40 $\Msun$.  For the intermediate 'case B'
yield models and the assumed solar ratio of iron to hydrogen of
$1.8\times 10^{-3}$ by mass \citep{anders_grevesse89}, we obtain
$\eta_{\rm Fe} \approx 0.6$.  This estimate is consistent with the
enhanced ratios $[\alpha/\mbox{Fe}] \approx 0.3$ observed for the
globular clusters in the Galaxy
\citep{carney96,lee_carney02,smith_etal02}, M87 \citep{cohen_etal98},
and M49 \citep{cohen_etal03}.  The conversion from the total
metallicity due to SNII to the iron abundance is $\mbox{[Fe/H]} =
\lg(\eta_{\rm Fe} Z_{\rm II} / Z_{\sun})$.

Table \ref{tab:gcc} compares the medians and the 25\% and 75\%
quartiles of the size and metallicity distributions of the model
clusters against the corresponding distributions for the Galactic GC
\citep{harris96}.  There is good agreement between our predictions
and the observations.  It is plausible that small discrepancies for
the smallest sizes and highest metallicities can be explained by
subsequent evolution of the cluster population at $z<3$, which we
discuss in \S\ref{sec:future}.  Note that the evolution can also
modify the quartiles of the distribution.  We therefore show in
Table~\ref{tab:gcc} both the quartiles for all Galactic clusters and
for the clusters with metallicities in the same range as in our model
(i.e., $[{\rm Fe/H}]<-1]$).  It is also interesting that in agreement
with observations no clusters are formed with very low (Pop III)
metallicities.

Note that at all epochs the dynamical time of the parent molecular
cores is very short ($\sim 10^6$~yr), which means that the galactic
gas is pre-enriched even before the first clusters form.  It follows
then that the oldest globular clusters do not contain the oldest stars
in the Galaxy.

The lack of the metal-rich clusters in our model compared to the
Galactic clusters (c.f. Table \ref{tab:gcc}) is likely to be explained
by the clusters forming in the higher-metallicity gas at $z<3$ (see
\S~\ref{sec:future}), not accounted for in our analysis.  It should be
noted, however, that a significant spread in metallicity distributions
exists for different galaxies \citep[e.g.,][]{harris01} and certain
differences with the simulated system are expected.  This issue can
be addressed in future studies by simulating globular cluster systems
in a number of galaxies.

\section{The Origin and Universality of the Globular Cluster Mass Function}
  \label{sec:mf}

One of the most important characteristics of globular cluster systems
is the mass function (GCMF).  Interestingly, the mass function derived
in the simulation is similar to the mass function of molecular clouds
in our and external galaxies, $dN/dM \propto M^{-\alpha}$ with $\alpha
= 1.4 - 2$ \citep[e.g.,][]{solomon_etal87,wilson_etal03}.  It is also
similar to the high-redshift mass function of dark matter halos in the
hierarchical CDM cosmology \citep{press_schechter74,sheth_tormen99}.
In this section we investigate the origin of GCMF in relation to the
mass function of giant molecular clouds and parent halos.

\cite{gnedin03} used a simple semi-analytic model in which a single
massive cluster dominates the mass of the globular cluster system
within a progenitor halo: $M_{\rm max} \lesssim M_{\rm GC} \approx
\varepsilon_{\rm GC}^{\rm b} f_{\rm b} M_{\rm h}$
(c.f. \S~\ref{sec:efficiency}).  The model implies that the shape of
the high-mass tail of the GCMF simply reflects the shape of the mass
function of progenitor halos of the Milky Way at high redshifts.  This
direct connection between the cluster and halo mass functions is due
to the assumption that GC properties are determined solely by the mass
of their parent halo.  This key assumption can be tested against the
results of our simulation.

We find indeed that the most massive cluster contributes a significant
fraction of the total cluster mass.  The average for all halos is
$M_{\rm max} \approx 0.6 \, M_{\rm GC}$.  In \S~\ref{sec:efficiency}
we have shown that $M_{\rm GC}$ is roughly proportional to the parent
galaxy mass, and therefore a similar relation exists for the most
massive clusters:
\begin{equation}
  M_{\rm max} = 2.9 \times 10^6 \, \Msun
     \left({M_{\rm h} \over 10^{11} \, \Msun}\right)^{1.29 \pm 0.12}.
  \label{eq:mmax}
\end{equation}
However, a significant scatter around this average relation is such
that for a given halo the masses of individual clusters can vary by a
factor of three.  Thus, the mass function of globular clusters does
not follow directly from the mass function of their parent halos.

\begin{figure}[t]
\plotone{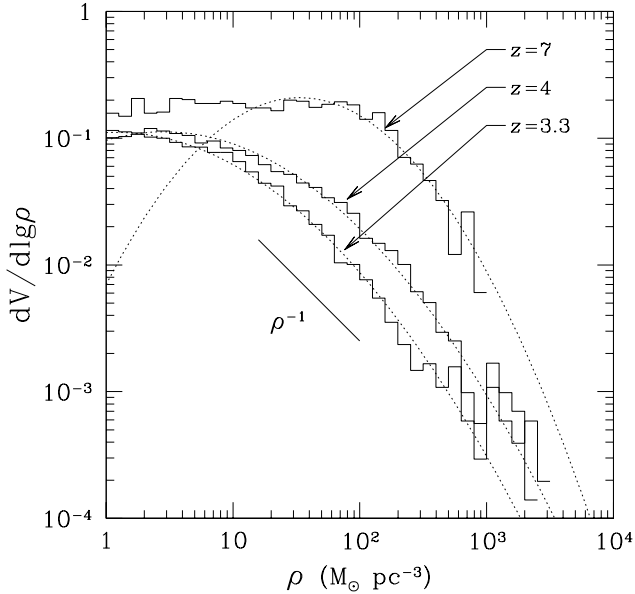}
\caption{Probability distribution functions of the gas density
  (fraction of the total volume occupied by the cells in a given density
  range) for the highest level of refinement at $z=7$, 4, and 3.3.
  The dotted lines
  show the log-normal fits to the highest-density tails
  of the distribution, in the range $1 - 10^3 \, \Msun$ pc$^{-3}$.
  The straight lines show the slopes $dV/d\lg{\rho} \propto \rho^{-1}$
  and $\rho^{-1.5}$.
  \label{fig:pdf}}
\end{figure}

Instead, we find that the overall shape of GCMF in our model is
determined by both the mass function of progenitor halos {\em and} the
mass function of molecular cloud cores within individual halos.  The
latter, in turn, is determined by the structure of the galaxy disks,
and in particular by the density probability distribution function
(PDF).

The shape of the PDF has been studied in several numerical simulations
of the turbulent ISM.  A log-normal distribution is thought to be a
generic feature of {\em isothermal} turbulent flows
\citep[e.g.,][]{vazquezsemadeni94,padoan_etal97,vazquezsemadeni_etal00}.
For non-isothermal supersonic flows \citet{scalo_etal98} found that
the PDF is a power law of density, although \citet{nordlund_padoan99}
argued that even in this case the PDF can be described equally well by
a power law or a log-normal function.  The log-normal shape is likely
to be due to the chaotic nature of the supersonic turbulent flows,
characterized by numerous random convergent flows and shocks.  The
evolution of individual gas elements can be thought of as a random
walk in density leading to the log-normal equilibrium distribution
\citep{elmegreen02}.

For the heating/cooling rates adopted in our simulation the gas at
densities $\rho_{\rm g}\gtrsim 5\ \rm M_{\odot}pc^{-3}$ cools
efficiently to the
lowest allowed temperature ($T_{\rm min} = 300$ K) and is therefore
nearly isothermal.  Accordingly, we expect the PDF in the simulation
to be log-normal.  The width of the log-normal PDF is proportional to
the rms Mach number of gas clouds \citep{padoan_etal97}, which
increases with time in the hierarchically assembled galaxies.  Thus we
expect the PDF to widen with decreasing redshift.

Figure \ref{fig:pdf} shows the density PDF measured in the simulation
at $z=7$, 4, and 3.3, using only the highest refinement level ($l=9$)
cells which cover the galactic disks and include the sites of GC
formation.  The log-normal distribution provides a very good fit to
the high density tail of PDF at $\rho > 1 \, \Msun$ pc$^{-3}$
\citep[see also][]{wada_norman01,kravtsov03}:
\begin{equation}
  {dN \over d\lg{\rho}} \propto 
     \exp{\left[-{(\lg{\rho}-\lg{\rho_0})^2 \over 2 \sigma_\rho^2}\right]}.
\end{equation}
The characteristic density $\rho_0$ and the dispersion $\sigma_\rho$
of the log-normal PDF vary with redshift. The characteristic density
decreases from $\lg\rho_0\approx 1.8$ at $z=8$ to $\lg\rho_0\approx 0.08$ at
$z=3.3$, while the width of the distribution $\sigma_{\rho}$ increases
from 0.46 to 0.85 at the same redshifts.  The evolution in this
redshift interval can be fit by $\lg{\rho_0} = 3.30 - (1.41 \pm 0.02)
[10/(1+z)]$ and $\sigma_\rho = 1.23 - (0.83 \pm 0.05)[(1+z)/10]$.

Figure \ref{fig:pdf} shows also that over a limited density range, $1
< \lg{\rho} < 3$, the PDF can be described by a power-law
\begin{equation}
  {dN \over d\lg{\rho}} \propto \rho^{-n}
  \label{eq:rho_pdf}
\end{equation}
with $n = 1.08 \pm 0.06$ (for $z=4$).  Over a wider range of
densities, however, the log-normal function is a somewhat better description of
the PDF.  For instance, the power-law slope becomes steeper with
increasing density: we find $n = 1.26 \pm 0.08$ for $\lg{\rho} > 1.5$
and $n = 1.41 \pm 0.13$ for $\lg{\rho} > 2$.  The latter range
includes the molecular cloud cores in which the massive clusters ($M >
10^5 \, \Msun$) form in our model. 

We should note that the molecular clouds in the simulation are only
marginally resolved.  Their temperature and structure depend on the
uncertain physics such as the presence of dust, cooling due to metals,
UV heating by nearby stars and radiative transfer. The form of the PDF
can depend on these properties and at this point we cannot reliably
differentiate between log-normal and power law mass functions.
However, the difference between the two at the interesting densities
is quite small and is not critical for our discussion.

Globular clusters in our model form only in the highest-density cores
of the identified molecular clouds which represent a subset of all
high-density cells.  Nevertheless, we find that the PDF of the cores
(or density peaks) is similar to the overall cell PDF at $\lg{\rho} > 0$.
The Kolmogorov-Smirnov test of the unbinned probability distributions
shows that the differences between the core PDF and the cell PDF are
not statistically significant.  At all epochs the probability that the
two PDFs are drawn from the same distribution is at least 13\%.  This
indicates that the density PDF of the cores is also described by the
same log-normal distribution.  The density of each core determines the
mass of the globular cluster it hosts.  The density PDF of the cores
thus determines the mass function of globular clusters.

Given our subgrid model, the cluster mass scales with the core gas
density as $M \propto \rho^{3/2}$ (eqs. [\ref{eq:m}] and [\ref{eq:r}]).
For a power-law
density PDF the expected cluster distribution is $dN/dM \propto
M^{-1-2n/3}$, or $\alpha = 1+2n/3$.  For $n=1$ this gives the slope
$\alpha = 5/3\approx 1.7$, while for $n=1.4\pm 0.13$, appropriate for
the highest-density tails of the PDF and the massive clusters,
$\alpha= 1.94\pm 0.09$, in good agreement with the mass function slope
seen in Figure~\ref{fig:gc_m}.  Therefore, we can expect a relatively
shallow, $\alpha \approx 1.7$, mass function for the small-mass
clusters forming in lower-density cores and the steep, $\alpha \approx
2$, mass function for massive globular clusters forming in the densest
regions of the disk.  The range of slopes derived in our model is in
close agreement with observations
\citep{elmegreen_efremov97,zhang_fall99,degrijs_etal03,anders_etal04}.

\begin{figure}[t]
\plotone{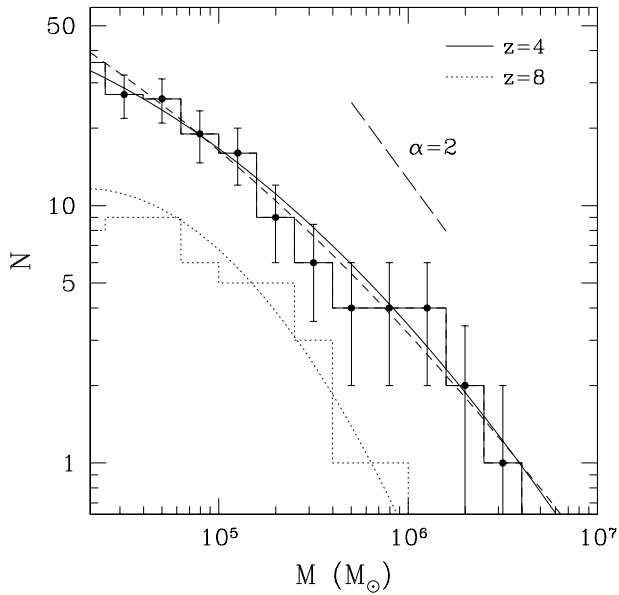}
\caption{Mass function of globular clusters forming at two epochs
  (histograms) with the superimposed fit (smooth solid and dashed lines)
  from the density PDF (eq. [\protect\ref{eq:mpdf}]).  The
  error bars represent Poisson errors $\propto N^{1/2}$.  The dashed line
  shows the mass function calculated from the density PDF of the lower,
  $l=8$, level of refinement normalized to the same total number of
  clusters.
  \label{fig:mpdf}}
\end{figure}

From our analysis above it is clear that the power-law is not a
unique description of the mass function.  The log-normal shape of the
density PDF implies the log-normal GCMF:
\begin{equation}
  {dN \over d\lg{M}} = N_0 \,
     \exp{\left[-{(\lg{M/M_0})^2 \over 2 \sigma_M^2}\right]}
  \label{eq:mpdf}
\end{equation}
with $\sigma_M = {3 \over 2} \sigma_\rho$ and $M_0 = M(\rho_0)$ using
equation (\ref{eq:m}).  Figure \ref{fig:mpdf} shows the mass
function of clusters at $z=8$ and $z=4$ along with the log-normal
function calculated from the best fit to the density PDF.  In other
words, the smooth lines in the figure represent not fits to the mass
function, but the fits to the PDF converted to the mass function using
our subgrid model. The log-normal mass function may thus be a reasonable
choice in fitting observed luminosity and mass functions of young clusters
when a simple power law does not provide a good fit. 

The parameters of the distribution, $M_0$ and $\sigma_M$, follow
directly from the parameters of the density PDF.  As redshift
decreases, the characteristic peak $\rho_0$ decreases and the
dispersion $\sigma_\rho$ increases.  While their exact values at a
given redshift may be specific to our simulated galaxy, i.e. depend on
the environment, the anti-correlation between the parameters may be
more general.  We find the following relation which can be tested by
future simulations and observations:
\begin{equation}
  \lg{M_0} = (6.2 \pm 0.5) - (2.8 \pm 0.5) \sigma_M.
  \label{eq:m0sig}
\end{equation}
These results apply to the mass function of young stellar clusters not
modified by dynamical evolution.

\begin{figure}[t]
\plotone{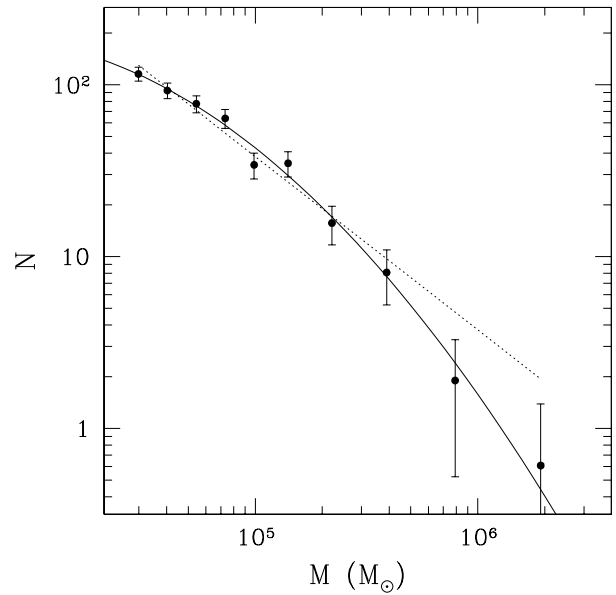}
\caption{The mass function of young (age $25<t<160$~Myr) star
  clusters (solid circles with error bars) in NGC 4038/9 (the
  ``Antennae''), as derived by \citet{zhang_fall99}, and the best fits of
  the power law (dotted line) and log-normal (solid curve) mass
  distributions.
  \label{fig:antennae}}
\end{figure}

In order to illustrate that the log-normal distribution can fit the
observations, we plot on Figure~\ref{fig:antennae} the mass function
of young (age $25<t<160$~Myr) stellar clusters in NGC 4038/9 (the
``Antennae''), as derived by \citet{zhang_fall99}.  This figure
shows that the log-normal function describes the observed mass
function as well as the power law.  The best fit power-law
slope is $\alpha=2.01 \pm 0.07$ with the reduced $\chi^2=1.17$, while
for the log-normal fit $\chi^2=0.68$.  The parameters of the fit, $\lg
M_0=3.7 \pm 0.4$ and $\sigma_{\rm M}=0.73 \pm 0.12$, are in reasonable
agreement with the derived relation (\ref{eq:m0sig}), especially if we
take into account the difference in redshifts, metallicity, and
environment.

The results presented in this section indicate that the observed
luminosity and mass functions of young clusters could be described by
a power-law due to the limited range of luminosities typically probed
in observations \citep[$\sim 2-4$ magnitudes or only about a factor of
$10-40$ in mass:][]{elmegreen_efremov97,whitmore00}.  If our results
are correct, the prediction would be that the power-law slope $\alpha$
should steepen at the highest cluster masses: $M \gtrsim 10^7 \,
\Msun$. Interestingly, the log-normal mass function implies that there
exists a maximum cluster mass at any given epoch.  This is an
important conclusion as it may explain the characteristic masses of
clusters in the Local Group.

\section{Numerical Convergence}
\label{sec:convergence}

The resolution of the simulation in any numerical study
invariably places constraints on the validity of the results.  The
spatial and mass resolution of the gas element in the adaptive mesh
simulation is set by the maximum level of refinement.  In our
simulation $l_{\rm max} = 9$ was determined by the available
computational resources.  We can check, however, how the results would
change if we limited the maximum level to $l=8$.  This corresponds to
factor of 8 lower mass resolution and a factor of two larger cells.

We find that the density PDF of the $l=8$ cells can be well fit by a
log-normal function but with a correspondingly lower characteristic
density $\rho_0$.  The dispersion $\sigma_\rho$ is consistent with
that for $l=9$ level within the errors.  The mass function resulting
from the $l=8$ density PDF is therefore somewhat steeper than for the
$l=9$ cells, because $M_0$ is lower and the same mass interval of
globular clusters falls on the steeper part of the log-normal
function.  The difference, however, is only apparent at very low
masses ($M \lesssim 3 \times 10^4 \ \Msun$).

The dashed line in Figure \ref{fig:mpdf} shows the cluster mass
function expected for the density PDF of the $l=8$ cells normalized to
the same number of clusters as the $l=9$ mass function.  It fits the
histogram of the model clusters equally well.  The deviation only
becomes significant at $M < 3 \times 10^4 \, \Msun$.  Thus, the shape
of the derived mass function converged at masses higher than $3 \times
10^4 \, \Msun$.

Our assumption that each molecular cloud forms only one cluster may
also affect the low mass end of the mass function.  If in reality the
cloud fragments into several self-gravitating cores, then smaller
clusters may form on the periphery of the cloud in addition to the
larger central cluster.  However, as we show in the next section,
these additional small clusters are likely to be quickly dissolved, so
that in the end the mass function of the surviving massive clusters
remains the same.

\section{Discussion}

\subsection{Evolution of Globular Clusters at lower redshifts}
  \label{sec:future}

In the preceding sections we analyze the formation of globular
clusters at $z \gtrsim 3$.  Although the limited computational
resources did not allow us to continue the simulation at the same
resolution to lower redshifts, in this section we conjecture on the possible
evolution of cluster population at later epochs.  

Analysis of our simulations hints that massive molecular clouds needed
for cluster formation are built in gaseous spiral arms. Their
formation may be enhanced in mergers between gas rich gas disks.  For
example, as we note in the caption of Figure 1, the gas disk shown in
this figure is actually in a state of very active accretion and
merging.  For instance, the two cold dense clumps clearly seen in the
temperature map are two satellite galaxies in the process of merging
with the disk.  The mass of the parent halo of the disk in Figure 1
increases by a factor of twenty between redshifts of 10 and 4 (a
period of only $\sim 1$~Gyr! See \S~2.2).

It is commonly thought that galaxy mergers create conditions conducive
to bursts of star formation and star cluster formation, as evidenced
by the starbursting galaxies \citep{kennicutt98,larsen02}.  In
particular, mergers stir and compress the interstellar gas creating
the high-pressure environments in which dense clouds and massive
stellar clusters can form \citep{elmegreen02}. Without the mergers,
star formation proceeds in a quiescent mode
\citep[e.g.,][]{abadi_etal02}.

The high rate of accretion and merging cannot be maintained at lower
redshifts. Statistically, the CDM halo merger rate at $z \lesssim 4$ decreases
rapidly as $(1+z)^{-2.5}$ \citep{gottlober_etal01}. In addition, as
galaxies evolve, most of the gas will be converted to stars so that
the large reservoir of gas needed to build up giant molecular clouds
may not be available.  Therefore, if galaxy mergers are connected to
globular cluster formation, a high merger rate between gas rich
galaxies at $z>3$ would lead to an almost continuous cluster
formation. At lower redshifts, on the other hand, mergers become rare
and the merging galaxies are gas deficient compared to their
high-redshift progenitors. Most of the subsequent formation of stellar
clusters may thus be limited to a single last major merger
event.  This, for example, could explain the bimodality of cluster
colors observed in many elliptical galaxies.

Finally, somewhat paradoxically, the globular clusters formed at high
redshifts may have a significantly higher chance of survival until the
present than clusters formed at later epochs.  The clusters in our
model form in extremely high-density environments within the galactic
disks.  Strong tidal forces in such regions are likely to disrupt
clusters quickly, unless they are located at the very center of the
parent galaxy or are ejected from the disk by a dynamical process
shortly after formation.  As only a fraction of clusters form in the
centers of progenitor galaxies, the latter mechanism should operate in
order for the old metal-poor clusters to survive until the present.

Frequent violent mergers at high redshifts may be just such a
mechanism.  Mergers disrupt the disks of the progenitor galaxies with
their existing stellar populations and impart a large amount of
orbital energy in the surviving clusters.  The high energy orbits
would allow globular clusters to spend most of the time in the
relatively low-density regions of the halo, outside the main disk.
Mergers could be responsible for a spheroidal distribution of the
globular cluster systems, even though the clusters actually form
within the disks of the progenitor systems. At low redshifts, on the
other hand, star clusters, even if they continue to form, may not be
able to escape the disk quickly enough to avoid disruption.

\begin{figure}[t]
\plotone{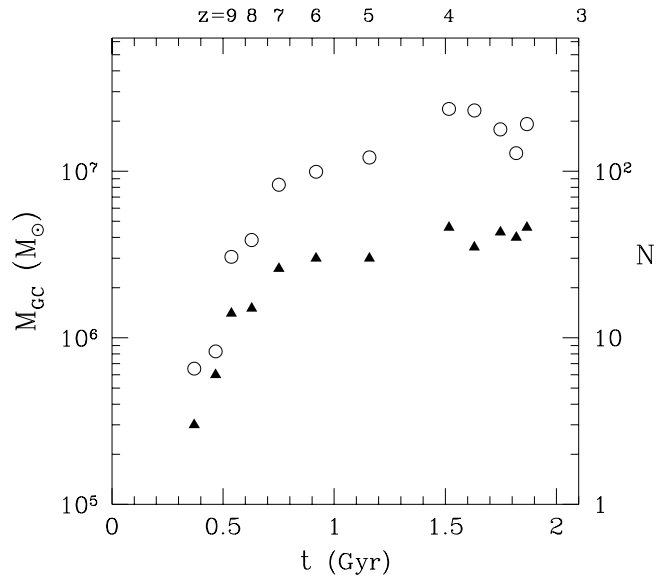}
\caption{The number of clusters (triangles) and
  the total mass in clusters (circles) formed at each simulation
  output according to our model.  The number of sites of cluster formation
  reaches a plateau at $z\approx 4$ and may decline at lower redshifts.
  \label{fig:time}}
\end{figure}

The effects discussed above may result in the preferred, albeit
extended, epoch of globular cluster formation at $z_{\rm GC}\sim 2-12$
($t_{\rm GC}\sim 0.3-3$~Gyr for the adopted cosmological model).  We
cannot prove this conjecture directly in our present simulations,
although one may argue that there are indications of this trend at
$z\sim 3-4$.  Figure \ref{fig:time} shows that the total mass of the
GC population and the number of clusters forming at a given epoch
increases until $z\approx 4$ and then saturate at a constant value at
lower redshifts.

Note that Figure \ref{fig:time} should be interpreted with caution.
Since clusters are not formed self-consistently during the simulation
run time but rather calculated a posteriori, it is impossible to
compute the rate of their formation.  In a self-consistent treatment,
clusters forming at a given epoch may exhaust the supply of gas in the
regions conducive to cluster formation and prevent the formation of
new clusters in the same place.  In our current analysis, on the other
hand, clusters at subsequent epochs can, in principle, form from the
same gas.  At the high redshifts that we considered, the gas
distribution changes sufficiently quickly and the rate of accretion of
new gas is very high, so that the epochs separated by tens of millions
of years can be safely considered independent.  This may not be true
for nearby epochs.  The flattening of $M_{GC}$ and $N_{GC}$ in
this figure can be interpreted as the fact that no new sites for
cluster formation are created.  Unfortunately, this is the best
estimate of the formation rate that we can provide with our current
simulation.  Our future work will address the formation of star
clusters at low redshifts directly.

To summarize, we conjecture that the formation of globular clusters at
$z\lesssim 2-3$ can be associated with the increasingly rare merger
events and would thus be progressively episodic.  The merger rates
derived from cosmological simulations \citep{gottlober_etal01}
indicate that on average galactic halos experience 1 to 4 major
mergers with the mass ratio $>0.2$ at $z<2$.  Two broad evolutionary
scenarios can be envisioned in this picture.  If the last major merger
occurs early ($z_{\rm merge} \gtrsim 2$), not too far from the
preferred epoch of GC formation, we would expect a continuous change
of the cluster properties.  If, on the other hand, the galaxy
experiences the last major merger late ($z_{\rm merge} \lesssim 1$),
with a substantial gap between $z_{\rm merge}$ and $z_{\rm GC}$, the
distribution of properties of the resulting cluster populations may be
bimodal.  The bi-modality in this case would reflect a considerable
change in the galactic environment (e.g., the metallicity of gas)
during the interval $\Delta z = z_{\rm GC}-z_{\rm merge}$.  This
latter scenario is particularly relevant to the formation of large
elliptical galaxies.

\subsection{Comparison with previous work}

Several recent studies explored the formation of globular clusters in
the context of hierarchical cosmology using both numerical simulations
and phenomenological semi-analytic models.  Here we discuss and
compare the specifics and results of our study to the previous similar
efforts.

\cite{weil_pudritz01} used a Tree-SPH simulation of the $\tau$CDM
($\Omega_0=1$) cosmology to study the large-scale distribution of
giant gas clouds at $z \lesssim 1$ in a $32 \, h^{-1}$ Mpc box.  The
authors analyzed collisions of baryonic clumps (clouds) in small-mass
dark matter halos within the context of the agglomeration model of
\citet{harris_pudritz94}.  They found a characteristic power-law
spectrum of cloud masses, $dN/dM_{\rm mc} \propto M_{\rm mc}^{-1.7}$,
similar to the mass function of molecular clouds and young star
clusters. The mass function of clouds and globular clusters in our
model is consistent with their result. The detailed comparison,
however, is not possible as the simulations of \cite{weil_pudritz01}
did not include star formation and the cooling of gas below $10^4$ K.
In addition, their low spatial resolution ($\sim 1$ kpc) prevented any
detailed study of the inner structure of the clouds and as well as the
density distribution and mass spectrum of clusters within individual
galaxies.

Recently, \cite{bromm_clarke02} used a Tree-SPH simulation to study the
collapse and fragmentation of gas during the evolution a single dwarf
galaxy $\sim 10^8 \ \Msun$ at $z \lesssim 24$.  The simulation assumed
that the gas was pre-enriched to the metallicity of $10^{-2} \,
Z_{\sun}$, which allowed the gas to cool to $\sim 1000$ K, and used
sink particles to follow the collapse of the gas in the highest
density ($n > 10^3$ cm$^{-3}$) regions.  The authors identified six
gas clumps with masses in the range $4 \times 10^4 - 2 \times 10^7 \
\Msun$, each associated with a separate small-mass dark matter halo.
They concluded that the characteristic mass of globular clusters is
determined by the characteristic mass of dark matter halos forming at
$z \gtrsim 10$.

The implicit assumption behind their conclusion is that the conditions
for cluster formation exist only at the earliest stages of galaxy
formation, prior to reionization.  The results of our study show that,
although the first globular clusters may form at $z>10$, the
conditions for cluster formation become more favorable at lower
redshifts when more gas accumulates in the disks of the progenitor
halos.  Reionization does not affect significantly the formation of
clusters in the relatively massive halos ($M_{\rm h} \gtrsim 10^{10} \
\Msun$).  Again, it is difficult to make a more detailed comparison,
given the very different setup of numerical simulations and physical
processes included.  Yet, it is worth noting that the gas density in
the most massive clump in the simulation of \cite{bromm_clarke02} (see
their Fig. 3) is well below the density of dark matter.  The average
gas density at the resolution limit in the inner 10 pc is only $\sim
10 \ \Msun$ pc$^{-3}$, lower than the observed density of globular
clusters (\S~\ref{sec:subgrid}).  The expected mass loss after cluster
formation would reduce the cluster density even further.  It is likely
that such clumps are still insufficiently dense to form real globular
clusters.

In contrast, the gas density at the sites of globular cluster
formation in our simulation is considerably higher ($\gtrsim 100 \
\Msun$ pc$^{-3}$) and is typically at least an order of magnitude
larger than the local density of dark matter.  We further assume that
the clusters form only at densities $\rho > 10^4 \ \Msun$ pc$^{-3}$
within the collapsing isothermal molecular cores.  The fact that GC
formation sites are strongly baryon-dominated explains the absence of
dark matter halos around globular clusters.  \cite{bromm_clarke02}, on
the other hand, conjecture that dark matter hosts of globular clusters
dissolve via violent relaxation before the cluster forms, while the
baryonic cores survive.

Both of the above studies attribute the shape of the GC mass function
to the distribution of their parent dark matter halos.  Our results
indicate that the situation is more complex.  We have shown that the
shape of the GCMF is determined {\em both} by the mass function of the
parent halos and the mass distribution of clusters within a single
halo (\S~\ref{sec:mf}).

\cite{beasley_etal02} extended the semi-analytical model of galaxy
formation {\tt GALFORM} to include a phenomenological prescription for
GC formation.  The model assumed that a constant fraction of stellar
mass, $\varepsilon$, would be in the form of globular clusters.  The
blue (metal-poor) clusters were associated with the quiescent mode of
star formation, while the red (metal-rich) clusters were assumed to
form during starbursts.  The parameters, $\varepsilon = 0.002$ for
blue clusters and $\varepsilon = 0.007$ for red clusters were set to
agree with the observed color distribution of the elliptical galaxy
NGC 4472.  However, \cite{beasley_etal02} truncated the formation of
blue clusters arbitrarily at $z = 5$ to create a distinctly bimodal
distribution of colors in their model.  In addition they find an
age-metallicity correlation for their model clusters, which is a
definite prediction of semi-analytical models.  In contrast, our
simulation follows the gas dynamics and metal enrichment
self-consistently and does not predict any clear age-metallicity
relation.

\subsection{Alternative subgrid models}
\label{sec:altern}

The results we presented are based on a particular subgrid model. The
main features of the adopted model are the assumption of the
isothermal cloud structure within the density peak of each molecular
cloud and the assumption that clusters form with a fixed efficiency at
densities above a constant density threshold (see
\S~\ref{sec:subgrid}).

Here we consider alternative subgrid models and show that they can
either be reduced to the model we use or cannot successfully reproduce
the mass function of globular clusters. Specifically, we consider the
following alternative functions for the cluster formation threshold:
(1) the Jeans mass, (2) the thermal pressure, and (3) the total
pressure contributed by thermal and by turbulent motions.  The size of
the star forming core, and therefore the cluster mass, is determined
differently by a threshold value of each of these functions.

In general, an alternative subgrid model can have a threshold
parameter that varies with the radius differently from the isothermal
gas density profile, $\rho_{\rm g} \propto r^{-2}$.  It is likely,
however, than in an isothermal cloud any function of interest would be
a power law, $f(r) \propto r^{-a}$.  The star formation radius
corresponding to the threshold value $f_{\rm csf}$ is then determined by
$f(R_{\rm csf}) = f_{\rm csf}$.

Since in the simulation we can only measure the parameter $f_{\rm
cell}$ averaged over the cell volume, we need to consider the volume
average
\begin{equation}
  f_{\rm av}(<r) = {1 \over V(r)} \int_0^r f(r) 4\pi r^2 dr
                 = {3 \over 3-a} f(r).
\end{equation}
The measured parameter is $f_{\rm cell} \equiv f_{\rm av}(<R_{\rm cell})$.
Therefore, the subgrid distribution is
\begin{equation}
  f(r) = {3-a \over 3} f_{\rm cell} \left({r \over R_{\rm cell}}\right)^{-a}.
\end{equation}
The radius of the star forming core is
\begin{equation}
  R_{\rm csf} = R_{\rm cell} 
            \left({3-a \over 3} {f_{\rm cell} \over f_{\rm csf}}\right)^{1/a}
  \label{eq:r_csf}
\end{equation}
and the enclosed mass is
\begin{equation}
  M(R_{\rm csf}) = M_{\rm cell}
            \left({3-a \over 3} {f_{\rm cell} \over f_{\rm csf}}\right)^{1/a},
  \label{eq:m_csf}
\end{equation}
where $M_{\rm cell} \equiv (4\pi/3) \rho_{\rm cell} R_{\rm cell}^3$ is
the amount of gas within a sphere embedded into the cell.  Including
the efficiency of star formation $\epsilon$, the stellar mass is $M_*
= \epsilon M(R_{\rm sf})$ and stellar radius is $R_* = R_{\rm
sf}/\epsilon$.

The Jeans mass is a critical mass of a cloud of density $\rho_g$ and
temperature $T$ that is unstable to gravitational instability, $M_{\rm J}
\propto T^{3/2} \rho_g^{-1/2}$.  For isothermal molecular clouds with
$T \approx\rm const$ and $\rho_g \propto r^{-2}$, a
threshold in $M_{\rm J}$ reduces to a corresponding threshold in $\rho_g$,
i.e. our original model.  Similarly, the thermal pressure criterion
$P_{\rm th} \propto \rho_g T$ reduces to the density criterion.

\begin{figure}[t]
\plotone{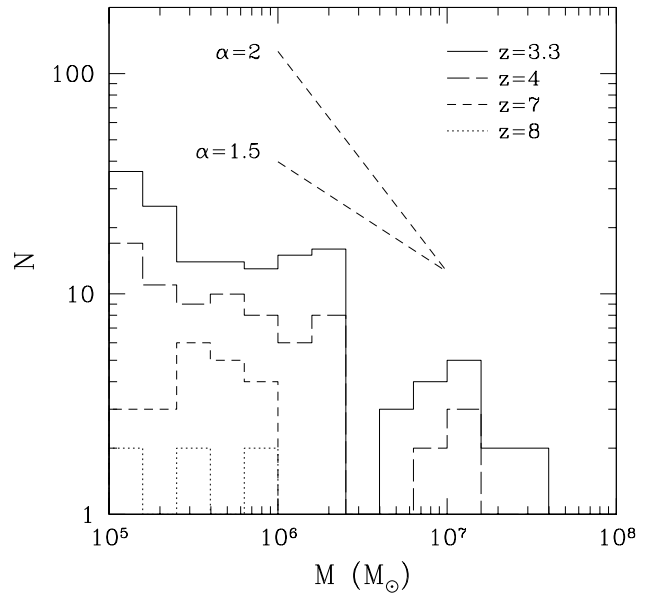}
\caption{The build-up of the initial mass function of globular clusters
  based on the turbulent pressure criterion.
  Dotted, dashed, long-dashed, and solid histograms show cumulative
  distributions at $z = 8$, 7, 4, and 3.3, respectively.  The
  straight dashed lines show two power-laws, $N\propto M^{-\alpha}$.
  Note that the mass function is significantly
  shallower than that with the density criterion, $\alpha = 2$.
  \label{fig:gc_mp}}
\end{figure}

The only variable independent of the gas density, at least in
principle, is the turbulent velocity dispersion, $\sigma$.  Turbulence
in the galactic disks is created by gravitational motions on scales
larger than a single cell and reflects large-scale flows around the
cloud.  In the simulation, the turbulent pressure, $P_{\rm turb} =
\rho_g \sigma^2$, typically dominates over the thermal by one or two
orders of magnitude, so that the total pressure $P \approx P_{\rm
turb}$.  The velocity dispersions and sizes of the Galactic giant
molecular clouds satisfy the following relation: $\sigma \propto
r^{1/2}$ \citep{larson81,brunt_heyer02}.  If the turbulent velocity
dispersion scales with the radius inside the cell as $\sigma \propto
r^{1/2}$, while $\rho_g \propto r^{-2}$, then $P \propto r^{-1}$.

A reasonable fiducial value for the threshold pressure, $P_{\rm csf}$,
can be obtained from observations of the Galactic globular clusters.
The median value of the pressure at the half-mass radius is $P_{\rm
obs} \sim 10^9$ K cm$^{-3}$.  The observable pressure in our model can
be derived taking into account the post-formation expansion of the
cluster.  Analogously to the observable density (eq. [\ref{eq:rho}]),
we find $P = \frac{3}{2} \epsilon^4 P_{\rm csf}$.  Setting $P = P_{\rm
obs}$, we obtain the star formation threshold $P_{\rm csf} = 5 \times
10^9$ K cm$^{-3} = 10^6 \ M_{\sun}$ pc$^{-3}$ km$^2$ s$^{-2}$.

The velocity dispersion in a cloud is calculated directly from the gas
velocities in the simulation.  First, the cloud is centered on the
cell of highest density.  Then, the center-of-mass motion, the net
radial motion, and the solid-body rotation velocity are subtracted
from the velocities of the cloud cells.  The remaining velocity field
is free of global organized motion.  The resulting velocity dispersion
is close to isotropic, as expected for a clean turbulent field.  We
use these turbulent dispersions, $\sigma$, to calculate the turbulent
pressure in the central cloud cell.  Then we apply equations
(\ref{eq:r_csf}) and (\ref{eq:m_csf}) to calculate the predicted
cluster sizes and masses, using $f_{\rm cell}=\rho_{\rm
cell}\sigma_{\rm cell}^2$, $f_{\rm csf}=P_{\rm csf}$, and $a=1$.

Figure \ref{fig:gc_mp} shows the cluster mass function calculated for
this subgrid model, which should be compared to Figure \ref{fig:gc_m}.
It is clear that the turbulent pressure criterion predicts a much
shallower mass function than that in the density threshold model.  The
power law slope is $\alpha \lesssim 1.5$. Such shallow slope is
inconsistent with observations. 

The value of the slope can be easily understood.  According to equation
(\ref{eq:m_csf}), the cluster mass is a certain fraction of the cell
mass.  This fraction depends on $(P_{\rm cell}/P_{\rm csf})^{1/a}$, and in
this case $a=1$.  If the fraction can be expressed as some power of
  the cell mass, $M_{\rm cell}$, then the cluster mass function can be
  derived from the mass function of central cloud cells (uniquely
  determined in the simulation), and the transformation would depend
  on the particular subgrid model.  Most of the densest cells are at
  the last refinement level and all have the same volume at any given
  epoch, so that $M_{\rm cell} \propto \rho_{\rm cell}$.  In agreement
  with the density PDF (eq. [\ref{eq:rho_pdf}]), the distribution of
  the central cells is roughly described by a power law
\begin{equation}
  {dN \over dM_{\rm cell}} \propto M_{\rm cell}^{-5/2}
  \label{eq:mf_mcell}
\end{equation}
for $M_{\rm cell} > 10^7 \ \Msun$, which are the cells harboring most
of the massive clusters.

Let us now derive the transformation from the cell mass to the cluster
mass.  The turbulent velocity does not in general scale with the cell
mass, since it is caused by large-scale flows, but we still find a
fairly robust correlation in the expected sense, $\sigma^2 \propto
M_{\rm cell}$, to within 0.3 dex in $\lg{\sigma^2}$.  Therefore, $M_*
\propto \rho_{\rm cell} P_{\rm cell} \propto M_{\rm cell}^3$.
Substituting this relation to equation (\ref{eq:mf_mcell}), we find
\begin{equation}
  {dN \over dM_*} \propto M_*^{-3/2}.
\end{equation}
This is comparable to the slope seen Figure \ref{fig:gc_mp} and is
significantly shallower than the observed slope.

The general formalism of calculating the cluster mass function
developed here gives us a powerful tool to evaluate the
alternative subgrid models.  Unless the transformation from the cell
mass to cluster mass is the same as for the density criterion ($M_*
\propto M_{\rm cell}^{3/2}$, leading to $\alpha \approx 2$), the
predicted cluster mass function would deviate from the observed.  The
criterion based on the turbulent pressure, at least in principle
independent of the density criterion, fails to reproduce the observed
mass function of young star clusters.

Of course, our prescription based on a simple threshold value of a
spherically-symmetric function does not capture all the details of the
molecular cloud physics, such as the filamentary internal structure,
dust, and magnetic fields.  These complications are beyond current
models of galaxy formation.  However, within the scope of our
spherically-symmetric approach to the molecular clouds, the cluster
formation criterion based on the density threshold seems to reproduce
the observations best.

\section{Summary}

We have presented a study of globular cluster formation at $z>3$ using
a very high-resolution cosmological simulation.  The clusters in our
model form in the high-density isothermal cores of giant molecular
clouds in dense gaseous disks of high-redshift galaxies.  The properties of
globular clusters are estimated using a simple physically-motivated
subgrid model.  Many of the observed properties of globular clusters
are reproduced naturally, without fine-tuning.  Our main
conclusions can be summarized as follows.

\begin{itemize}
  
\item Our results indicate that the first globular clusters in the
  Galaxy should have formed around $z \approx 12$ and cluster
  formation continued at least until $z \approx 3$.

\item 
  Most globular clusters in our model form in halos of mass $\gtrsim
  10^9 \ \Msun$.  The spatial distribution of these halos at $z>3$ is
  highly clustered (biased) with respect to the overall distribution
  of matter (\S~\ref{sec:spatial}).  The high spatial bias of their
  parent halos explains the present more concentrated radial
  distribution of globular clusters relative to dark matter. 
  
\item 
  Within the progenitor systems, globular clusters form in the
  highest-density regions of the disk: the cores of molecular clouds.
  In the most massive disk in the simulation, the newly formed
  clusters trace the spiral arms and the nucleus similarly to the
  young stellar clusters observed in merging and starbursting
  galaxies. 
  
\item 
  The mass function of globular clusters at birth can be approximated
  by a power-law $dN/dM \propto M^{-\alpha}$ with $\alpha \approx
  2$, in good agreement with observations of young star clusters
  in the Antennae.  It can also be well approximated by the log-normal
  function. The shape of the mass function is
  determined both by the mass function of parent galaxies and the mass
  distribution of molecular cloud cores within each halo.  Although
  the physical processes governing the evolution of the halos and
  molecular clouds are completely different, their mass distributions
  can both be approximated by the power-law functions with $\alpha
  \approx 1.5-2$.

\item 
  The halo mass function arises during the hierarchical build-up of
  structures in the expanding universe from the initial random
  perturbations.  The mass function of the molecular clouds cores, on
  the other hand, samples the underlying density
  probability distribution function (PDF) of the gas in the galactic
  disks. 

\item 
  The local efficiency of globular cluster formation within the parent
  molecular cloud is $M/M_{\rm mc} \approx 10^{-3}$, with 
  considerable scatter.  The global efficiency, the ratio of the total
  globular cluster mass to the baryonic mass of the parent galaxy, is
  $M_{\rm GC}/M_{\rm b} \approx (2-3) \times 10^{-4}$.  The mass in
  globular clusters scales with the total galaxy mass as $M_{\rm GC}
  \propto M_{\rm h}^{1.1}$ (eq. [\ref{eq:msum}]).  These values for
  the formation efficiencies are in general agreement with
  observations \citep{harris_pudritz94,mclaughlin99}.
  
\item We find that the mass of the globular cluster population and the
  maximum cluster mass in a given region strongly correlate with the
  local average star formation rate density: $M_{\rm max}\propto
  \Sigma_{\rm SFR}^{0.54\pm 0.07}$ and $M_{\rm GC}\propto \Sigma_{\rm
    SFR}^{0.75\pm 0.06}$ at $z=3.3$.  A similar correlation exists for
  the observed nearby galaxies \citep{larsen02}.  The correlation
  arises because both the star formation rate and mass of the globular
  cluster population are controlled by the amount of gas in the
  densest regions of the ISM. However, it is not clear whether this
  correlation is general or applies only to gas rich starbursting
  environments.

\item 
  Our model predicts the lack of clear age-metallicity and
  mass-metallicity correlations, at least for the clusters with $[{\rm
  Fe/H}] \lesssim -1$.  Although there is an overall trend of
  increasing the average metallicity with time, a significant spread
  of metallicities exists among different progenitor galaxies and
  within the interstellar medium of each galaxy.  The field stars and
  stellar clusters forming at a given epoch in the simulation exhibit
  scatter in metallicity of up to two orders of magnitude.

\item 
  The distribution of sizes and metallicities of the massive ($M >
  10^5 \ \Msun$) globular clusters match those of the Galactic
  globulars, with the exception of the largest size and the highest
  metallicity tail.  It is plausible that these discrepancies can be
  rectified by dynamical evolution and the continuing formation of
  globular clusters at $z<3$.

\end{itemize}

The simulation results presented in this paper, directly address only
formation of clusters at $z>3$.  We conjecture (\S~\ref{sec:future})
that the overall efficiency of globular cluster formation and chances
for their survival are significantly reduced at lower redshifts with
most clusters at $z\lesssim 2$ forming in rare gas-rich galactic
mergers. Although we cannot prove this with our current simulations,
we argue that the preferred formation epoch of globular clusters
surviving until the present is $z \sim 3-5$ (or $t \sim 1-2$~Gyr in
the adopted cosmology) when the gas supply is abundant in the disks of
the progenitor halos and the merger rate of the progenitors is high.
All old globular clusters thus would appear to have similar ages.

Although the high-redshift globular clusters form in dense {\it
  gaseous} disks, most the high-$z$ disks are expected to merge with
the Milky Way and disrupt by $z=0$.  The disrupted systems form
diffuse dark matter halo and contribute to the stellar halo of the
host. Their clusters would share the fate of the stripped stars of the
disrupted galaxies that build up galactic stellar halo and should
therefore have present-day spatial distribution similar to that of the
stellar halo.

In future work, we plan to study the details of the dynamical
evolution of the globular cluster population by extending our current
simulation to $z=0$.  We will incorporate our subgrid model of star
cluster formation directly in the high-resolution simulations of
galaxy formation.

Our present results are encouraging and demonstrate that globular
clusters with properties similar to the observed clusters form
naturally within young $z \gtrsim 3$ galaxy disks in the standard
$\Lambda$CDM cosmology.  The conditions for cluster formation
appear to be widespread at $z\sim 5$, coinciding with the peak of global star
formation rate.  As the formation of first stars marks the end of
cosmic dark ages \citep{rees97}, the formation of globular clusters
marks a veritable stellar renaissance of the Universe.

\acknowledgements 

We would like to thank L. Blitz, S. Boldyrev, M. Fall, D. Lamb, G. Meylan,
E. Ostriker, J. Truran, S. van den Bergh, and S. Zepf for stimulating
discussions on globular cluster formation, and Q. Zhang for providing
the mass function for the Antennae.  
We are also grateful to Nicole Papa for careful reading of the manuscript.
The simulations and analyses
presented here were performed on the IBM RS/6000 SP system at the
National Energy Research Scientific Computing Center (NERSC) and on
the Origin2000 at the National Center for Supercomputing Applications
(NCSA).  This work was supported by the National Science Foundation
under grants No. AST-0206216 and AST-0239759 (CAREER) to the
University of Chicago.  OYG is supported by the STScI Fellowship.

\bibliography{gc}

\end{document}